\documentclass[%
 reprint,
 amsmath,amssymb,
 aps,
 prapplied
]{revtex4-2}

\usepackage{hyperref}
\usepackage{multirow}
\usepackage{xcolor}
\usepackage{amsmath}
\usepackage{graphicx}
\usepackage{dcolumn}
\usepackage{bm}
\usepackage{datetime}
\newdateformat{monthyeardate}{\shortmonthname[\THEMONTH] \THEYEAR}

\usepackage{appendix}

\usepackage{fancyhdr}

\begin{document}

\preprint{APS/123-QED}

\title{Chiral Terahertz Amplification and Lasing \\ using Two-Dimensional Materials with Berry Curvature Dipole
}

\author{Amin Hakimi$^1$}
\email{amin.hakimi@uci.edu}

\author{J. Sebastián Gómez-Díaz$^2$}
\email{jsgomez@ucdavis.edu}

\author{Filippo Capolino$^1$}
\email{f.capolino@uci.edu}

\affiliation{$^1$ \mbox{Department of Electrical Engineering and Computer Science, University of California, Irvine, CA 92617, USA}\\
$^2$ \mbox{Department of Electrical and Computer Engineering, University of California, Davis, CA 95616, USA}}



 
\begin{abstract}
Compact, electrically driven sources of coherent terahertz (THz) radiation remain a challenge due to the lack of efficient gain media and scalable device platforms. Here, we propose and theoretically investigate a cavity-based THz gain mechanism enabled by Berry curvature dipole (BCD) in a DC-biased, low-symmetry two-dimensional (2D) material. Placing the biased 2D layer at the center of a Fabry–Pérot cavity enhances light–matter interactions, enabling direct conversion of DC electrical power into coherent THz radiation. We analyze the conditions for amplification and lasing, and identify the parameter regimes that support self-oscillatory coherent emission. Rather than introducing a specific device implementation, our work establishes the physical principles and operating conditions for BCD-enabled THz gain and lasing and provides the theoretical foundation for future realizations.
The chiral nature of BCD-induced response enables bias-tunable chiral optical gain, selective polarization eigenstate amplification, and electrically controlled handedness of the emitted radiation. Importantly, substantial amplification and lasing are achieved using only a single 2D material, significantly simplifying device design while preserving scalability across the THz band via cavity-length tuning. This platform is broadly applicable to low-symmetry 2D materials with finite BCD, offering a general route toward compact, frequency-tunable, and polarization-selective THz sources.

\textit{keywords}: Terahertz photonics, Berry curvature dipole, Two-dimensional material, Electro-optic effect, Light-matter interaction
\end{abstract}

\maketitle

\section{Introduction}

Efficient and miniaturized sources of coherent radiation in the THz frequency range (0.1-10 THz) remain a critical challenge in modern photonics due to the so-called “THz gap”, stemming from the lack of natural gain media and efficient radiation mechanisms in this frequency band. 
Bridging this gap is crucial, as THz waves enable non-ionizing imaging for medical, security, and industrial applications, support sensitive molecular spectroscopy, and promise high-speed, short-range wireless communication beyond conventional microwave bandwidths \cite{davies2002development,tonouchi2007cutting,pawar2013terahertz,nagatsuma2016advances,leitenstorfer20232023}.
Currently, THz generation relies on electronic, photonic, and quantum-based approaches. Electronic and optical methods typically use inefficient indirect mechanisms such as frequency mixing or difference-frequency generation \cite{tymchenko2017highly} while semiconductor quantum devices enable direct emission, as the notable case of quantum cascade lasers (QCLs), operating primarily in the 1-5 THz range \cite{gao2023recent}. Despite their success, QCLs face intrinsic constraints including cryogenic operation, limited tunability, and millimeter-scale cavities \cite{kainz2018barrier,young2009wavelength}. Consequently, the realization of room-temperature, broadband, coherent, and compact THz sources remains an open challenge \cite{lee2009principles,zhang2010introduction,zhang2021intense}.
Two-dimensional (2D) materials have recently offered a complementary approach to these conventional techniques, as their strong and unusual light-matter interactions and gate-tunable electronic properties open unprecedented possibilities for the development of compact, tunable THz modulators, detectors, and emitters that will conform the next-generation of THz technologies \cite{tassin2013graphene,wang2018thz,shi2018thz,cong2021coherent,qiu2021photodetectors,mittendorff20212d,wang2023giant,lin2023recent,pakniyat2024magnet,wang2023giant,qiushl25, de2025roadmap}.

\begin{figure}
    \centering
    \includegraphics[width=0.95\linewidth]{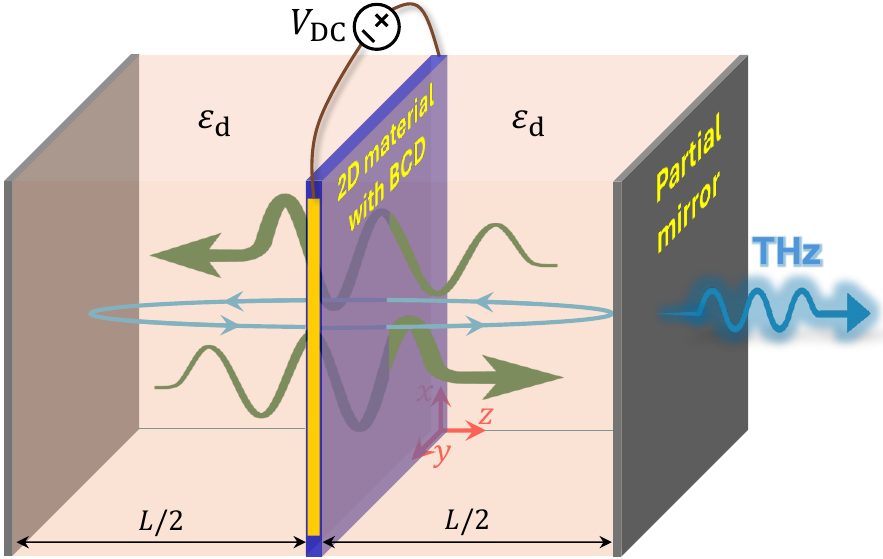}
    \caption{Schematic of a cavity enhancing light-matter interactions in a DC-biased 2D material exhibiting BCD. The cavity is closed at the right end by a partially reflecting mirror \textcolor{black}{through which light is emitted.}}
    \label{fig:Main}
\end{figure}

In a related context, the geometry of electronic wavefunctions in solids, encoded in the Berry curvature, underpins many topological and nonlinear phenomena \cite{xiao2010berry,nagaosa2010anomalous,bansil2016colloquium}. While often linked to time-reversal symmetry breaking, Berry curvature can be finite in nonmagnetic materials lacking inversion symmetry \cite{xiao2010berry}. An asymmetric distribution of Berry curvature around the Fermi surface gives rise to the Berry curvature dipole (BCD), the first moment of the Berry curvature over occupied states \cite{sodemann2015quantum}, which signals geometric deformation in electronic bands and drives nonlinear quantum effects in time-reversal-invariant systems. Materials with a finite BCD exhibit rich nonlinear electronic and optical responses, including the nonlinear Hall effect \cite{du2018band,ma2019observation}, current rectification \cite{kumar2021room,sinha2022berry}, kinetic magnetoelectric effects \cite{Furukawa2017,Calavalle2022}, and optical phenomena such as the circular photogalvanic and kinetic Faraday effects \cite{Xu2018,tsirkin2018gyrotropic}, and enhanced second-harmonic generation \cite{fu2023berry}. Recent studies suggest that in low-symmetry 2D materials and non-centrosymmetric bulk materials, the BCD induces a non-Hermitian electro-optic (EO) response that generates a nonreciprocal optical gain dependent on light polarization and propagation direction \cite{lannebere2022nonreciprocal,Shi2023,rappoport2023engineering,hakimi2024chiral,morgado2024non,eslami2025berry,lannebere2025symmetry}. Platforms such as twisted-bilayer graphene (TBG) \cite{rappoport2023engineering,hakimi2024chiral}, WTe$_2$ \cite{Xu2018,Kang2023}, twisted double bilayer graphene \cite{de2025metallic}, and Weyl semimetals \cite{zhang2018berry} are promising due to their large, tunable BCDs. Moreover, robust band structures can maintain large BCD even at room temperature \cite{nishijima2023ferroic}. Unlike conventional amplification relying on population inversion, this mechanism arises from intraband Bloch-electron dynamics under static fields \cite{roldan2025optical, eslami2025berry}, where the optical wave drives electrons to emit additional in-phase radiation, conceptually akin to free-electron laser amplification \cite{o2001free}.

Enhancing light-matter interactions strengthens the coupling between electromagnetic fields and material excitations. Engineered field enhancement via cavities, metasurfaces, or plasmonic structures has led to novel quantum and optoelectronic devices \cite{ramezani2016plasmon,flick2018strong,forn2019ultrastrong,berghuis2020light,gonccalves2020strong,tao2021enhancing,jo2023direct}.
Here, we propose a cavity-based platform incorporating a single 2D material with BCD at its center to efficiently transfer energy from a DC bias into THz resonant modes (Fig. \ref{fig:Main}). This DC-driven Fabry-Pérot (FP) cavity drastically enhances light-matter interactions on the 2D layer, \textcolor{black}{compared with the free-space configuration considered in \cite{rappoport2023engineering}, thereby} increasing the overall THz amplification \textcolor{black}{and lasing} and enabling a frequency-tunable and scalable operation across the THz range. The chiral nature of the BCD-induced gain locks the handedness of the emitted light to the sign of the applied bias \cite{hakimi2024chiral,lannebere2025chiral}. Contrary to devices based on stacks of 2D materials \cite{hakimi2024chiral}, the proposed platform relies on a single 2D layer, which greatly simplifies its design and fabrication process and facilitates future experimental demonstrations. \textcolor{black}{Moreover, here, the lasing frequency is primarily determined by the cavity length rather than the combination of length and wavelength-frequency dispersion as in \cite{hakimi2024chiral}, providing greater flexibility in designing the cavity to operate at a desired frequency. Finally, despite employing only a single 2D material, the DC bias required to reach the lasing threshold remains comparable to that reported in previous work using a stack of 2D materials \cite{hakimi2024chiral}.}
Rooted in a semi-classical Maxwellian formalism, we demonstrate that (i) THz amplification can be obtained and engineered even in the case of 2D layers with relatively large scattering rates; and (ii) damping mechanisms can be effectively mitigated by increasing the cavity quality factor, raising the DC bias, or operating at higher-order longitudinal cavity resonances. We further analyze the platform's self-oscillatory lasing condition using complex-frequency analysis, identifying both the lasing threshold and the oscillatory frequency. Finally, we derive an analytical expression for the amplification threshold under high-frequency (low-loss) and low-frequency (high-loss) approximations, providing a practical guide to estimate the DC bias required to achieve THz amplification in practice. Cavity-based BCD platforms offer a practical route for tunable, scalable, and handedness-selective THz amplification and lasing.

\section{Cavity Mediated Amplification}
\label{sec:T_and_R}

\begin{figure}[t]
    \centering
    \includegraphics[width=1\linewidth]{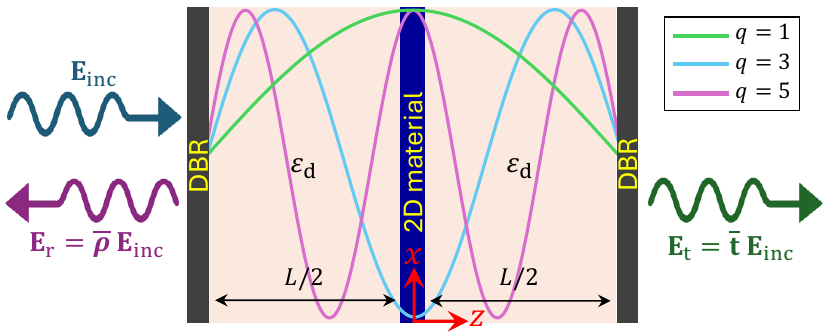}
    \caption{Schematic of a Fabry-Pérot cavity enhancing light-matter interactions with a 2D material with BCD positioned at its center, and distributed Bragg reflectors (DBRs) placed at both ends acting as partially reflective mirrors. The longitudinal electric field profiles for odd-order resonances ($q=1,3,5$) are shown, highlighting a field maximum at the cavity center. The 2D material with BCD is DC-biased, generating a transverse drift current resulting in optical chiral gain, with amplification controlled by the applied DC bias.}
    \label{fig:Structure}
\end{figure}

In an ideal FP cavity with perfectly reflecting mirrors (i.e., reflection coefficient magnitude equal to unity), the allowed wavenumbers of the longitudinal cavity modes are $k_q = q\pi/L$, where $q=1,2,3, \dots$ denotes the longitudinal mode index and $L$ is the separation distance between the mirrors. For a cavity oriented along the $z$ direction \cite{saleh2019fundamentals}, the corresponding longitudinal electric field mode profiles take the form $\sin\left(k_q \left(z+L/2 \right)\right)$ for $-L/2 \le z \le L/2$, as illustrated in Fig.~\ref{fig:Structure}. These profiles exhibit alternating symmetry with respect to the cavity center $(z=0)$: mode profiles with odd $q$ are symmetric and reach a maximum magnitude at the cavity center, whereas mode profiles with even $q$ are antisymmetric and have a node (zero) at the cavity center.

Placing a 2D material capable of optical gain, such as TBG \cite{rappoport2023engineering,hakimi2024chiral}, at the cavity center $(z=0)$ maximizes light-matter interactions and enables amplification of the odd cavity modes when sufficient gain is provided. Specifically, we propose the structure illustrated in Fig. \ref{fig:Structure}, in which the 2D material is positioned at the center of the cavity to strongly interact with odd-order modes, and distributed Bragg reflectors (DBRs) are placed at both ends to serve as partially reflective mirrors. By adjusting the cavity length, the supported odd modes can be engineered within the frequency range where TBG provides maximum amplification (0.1–4 THz), thereby achieving the highest possible amplification in the transmitted wave.

In the following, we first analyze the polarization eigenstates of the 2D material and the associated dissipated power, deriving the conditions under which the damping becomes negative, indicating amplification. Next, we summarize the method to explore the transmittance and reflectance spectra of the cavity platform when it is excited by right- and left-circularly polarized (RCP and LCP) waves. Finally, we evaluate amplification, gain, and the potential for lasing using a single TBG layer embedded in the FP cavity and examine the platform performance metrics versus the magnitude and sign of the BCD gain parameter, $\xi$.

\subsection{Polarization Eigenstates and Gain Condition in 2D Material}

Although the concepts in this work are general for any 2D material with BCD, we specifically consider TBG, whose conductivity matrix in the $x,y$ basis is expressed as \cite{rappoport2023engineering}

\begin{equation}
\overline{\boldsymbol{\sigma}}(\omega)= \left[\begin{array}{cc}
\sigma_{xx} & \sigma_{xy} \\
\sigma_{yx} & \sigma_{xx}
\end{array}\right] =
 \frac{\sigma_0}{\gamma-i \omega}\left[\begin{array}{cc}
\omega_{\mathrm{F}} & \xi \\
0 & \omega_{\mathrm{F}}
\end{array}\right]-\frac{\sigma_0}{\gamma}\left[\begin{array}{cc}
0 & -\xi \\
\xi & 0
\end{array}\right],
\label{eq:TBG}
\end{equation}
where $\sigma_0 = 2e^2/h$ is the conductance quantum, $\gamma = 1/\tau$ is the scattering rate, and $\omega_{\mathrm{F}} = E_{\mathrm{F}}/\hbar$ , with $E_{\mathrm{F}}$ denoting the Fermi level. The parameter $\xi$ controls gain and rules the linear EO response; it is proportional to the applied static electric field $E_0$ and to the BCD of the material $D_{\rm B}$: $\xi = \pi e D_{\rm B}E_0/\hbar$, as described in \cite{rappoport2023engineering,hakimi2024chiral}. Note that $\xi$ is a real-valued parameter, since both the applied DC field and $D_{\rm B}$ are real. However, depending on the orientation of the DC bias, $\xi$ can take either positive or negative values. \textcolor{black}{In the absence of the applied DC bias, there is no EO response.}

\textcolor{black}{The conductivity model in Eq. \eqref{eq:TBG} is obtained by considering a static electric field $E_0$ applied along the $y$-axis, as illustrated in Fig. \ref{fig:Main}, and evaluating the current-density response to a weak oscillatory electric field around the biasing point. In this model, the interband contribution to the optical conductivity is neglected. Moreover, higher-order terms contributing to the linear EO conductivity vanish for time-reversal-symmetric (TRS) 2D materials. For 2D materials with broken TRS, these higher-order terms are proportional to $\tau^2$ and therefore provide much smaller contributions than the first-order contribution included in Eq. \eqref{eq:TBG} \cite{rappoport2023engineering}.
Additionally, the BCD components of a 2D material are defined as the momentum-space integral 
\begin{equation}
    \left(D_{\rm B}\right)_{i,j} = - \frac{1}{\left(2\pi \right)^2} \int \frac{\partial f_0}{\partial{k_i}} \Omega_{\mathbf{k},j} d^2\mathbf{k},
\end{equation}
where $f_0$ is the equilibrium Fermi-Dirac distribution and $\Omega_{\mathbf{k}}$ is the Berry curvature in momentum space \cite{sodemann2015quantum}. Thus, although the BCD is determined by the momentum-space distributions of Berry curvature and carrier distribution, it is itself an integrated material parameter rather than a momentum-dependent quantity. Consequently, $D_{\rm B}$, and hence $\xi$, is treated as a local material parameter in the conductivity model, which is valid for propagating waves. For 2D materials, $D_{\rm B}$ has units of length (e.g., nm), and therefore $\xi$ has the unit of ${\rm s}^{-1}$.}

The \textcolor{black}{anisotropic} conductivity model described in Eq. \eqref{eq:TBG} consists of two contributions. The first matrix represents the non-Hermitian component (i.e., neither purely Hermitian nor anti-Hermitian), which accounts for energy exchange between the 2D material and the interacting wave and can therefore lead to optical gain or loss. The second matrix represents the gyrotropic component of the conductivity, which is anti-Hermitian (skew-Hermitian), i.e., not associated with gain or loss, and is responsible for inducing nonreciprocity between the two polarization eigenstates of the electric field. The resulting eigenvalues of the conductivity matrix are 
\begin{equation}
\sigma_{1,2} = \sigma_{xx} \mp \sqrt{\sigma_{xy} \sigma_{yx}},
\label{eq:eigenval}
\end{equation}
and the corresponding polarization eigenstates (eigenvectors) of the 2D material's conductivity are \cite{hakimi2024chiral}

\begin{equation}
\mathbf{E}^{\sigma}_{1,2}=
\left[\begin{array}{cc}
\mp \sqrt{\frac{\sigma_{xy}}{\sigma_{yx}}} \\
1
\end{array}\right] =
\left[\begin{array}{cc}
\mp i\sqrt{\frac{\omega+i2\gamma}{\omega+i\gamma}} \\
1
\end{array}\right].
\label{eq:eigenvectors}
\end{equation}
As discussed in \cite{hakimi2024chiral}, for a positive gain parameter $\xi>0$, the eigenvalue $\sigma_2$ corresponds to a lossy polarization eigenstate, while $\sigma_1$ leads to gain when ${\rm Re}(\sigma_1)<0$. The amount of gain is controlled by the parameter $\xi$, which is responsible for the linear EO response. \textcolor{black}{In general, the polarization eigenstates have elliptical polarizations that become predominantly circular at intermediate and high frequencies, particularly in the limit $\omega \gg \gamma$. At low frequencies or for large loss, corresponding to a non-small $\gamma/\omega$, the eigenstates become more elliptical. Indeed, as demonstrated in \cite{rappoport2023engineering}, the incident polarization that maximizes amplification is not necessarily circularly polarized. However, optimizing the incident polarization provides only a slight improvement in amplification compared with circular polarization, with a maximum enhancement of less than 5\% in the considered parameter range, whereas varying $\xi$ can lead to a much more substantial enhancement. Therefore, we employ circularly polarized waves as the incident excitation of the cavity (Fig.~\ref{fig:Structure}), providing strong coupling to the predominantly circular polarization eigenstates while avoiding the need to optimize the incident polarization at each frequency.} Moreover, due to the chiral nature of the BCD–induced gain, the polarization eigenstate that exhibits gain can be selected by the sign of the parameter $\xi$. This results in chiral modes, which will be discussed in more detail in the context of the scattering and modal analysis of a cavity incorporating this 2D material. Notably, both the non-Hermitian and gyrotropic contributions to the conductivity are required to produce two distinct polarization eigenstates with nonreciprocal propagation: one experiencing gain and the other loss. Therefore, for positive $\xi$, the polarization eigenstate $\mathbf{E}^{\sigma}_{1}$ is the one that can experience amplification, with RCP handedness when propagating along the $+\hat{\mathbf{z}}$ direction and the opposite handedness when propagating along the $-\hat{\mathbf{z}}$ direction.

To estimate the magnitude of the $\xi$ required for amplification in the 2D material via the amplifying polarization eigenstate $\mathbf{E}^{\sigma}_{1}$, we examine the power dissipated by the conductivity eigenvalue $\sigma_1$. The surface power density dissipated in the 2D material is given by \cite{Felsen1973radiation}

\begin{equation}
    p_{\rm dis} = \frac{1}{2} {\rm Re}\left\{ \mathbf{E}^* \cdot \mathbf{J} \right\} = \frac{1}{2} \left| \mathbf{E}_1^{\sigma} \right|^2 {\rm Re}\left(\sigma_1 \right).
    \label{eq:power}
\end{equation}
When ${\rm Re}\left(\sigma_1 \right) < 0$, the dissipated power density is negative, indicating energy transfer {\em from} the material {\em to} the optical wave that is the onset of amplification provided by the 2D material.
\textcolor{black}{To establish lasing in Section \ref{sec:lasing}, not only power needs to be amplified, but it also has to be amplified more than the total (intrinsic and external) loss experienced by a cavity.}

A simple and practical estimate of the $\xi$ value required for amplification is based on a Taylor-series expansion, as detailed in the Supporting Information, considering two regimes determined by the relative magnitude of the angular frequency $\omega$ and the loss parameter $\gamma$, yielding

\begin{equation}
    \begin{cases}
      \xi/\omega_{\rm F} > 2\gamma/\omega, \quad & \omega \gg \gamma\\
     \xi/\omega_{\rm F} > \gamma/(2\sqrt{2}\omega), & \omega \ll \gamma .
    \end{cases}
\label{eq:gain_condition}
\end{equation}
These are necessary conditions for amplification. However, the amount of amplification is determined by the magnitude of the field $\mathbf{E}^{\sigma}_{1}$ at the center of the cavity, hence by the cavity's quality factor.

\subsection{Transfer Matrix Formulation}
\begin{figure}[t]
	\centering
	\includegraphics[width=0.85\linewidth]{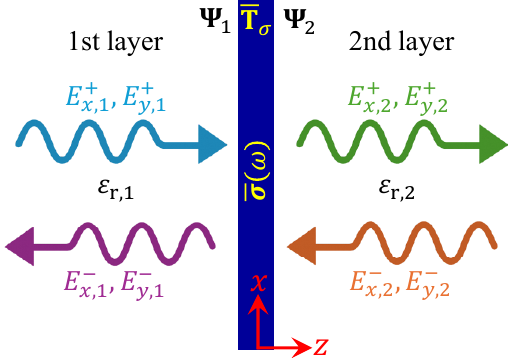}
	\caption{Geometry of the structure used in the transfer-matrix derivation. A 2D material with surface conductivity $\overline{\boldsymbol{\sigma}}(\omega)$ is sandwiched between two dielectric layers with different dielectric constants. Forward- and backward-propagating waves are shown on both sides of the 2D material.}
	\label{fig:MultiLayer}
\end{figure}

The transfer matrix method (TMM) provides a systematic framework to analyze the electromagnetic response of multilayered structures, and is especially useful for studying wave propagation, reflectance, transmittance, and absorptance in both periodic and non-periodic multilayer systems \cite{hao2008electromagnetic}. We begin by deriving the transfer matrix of a 2D material described by an anisotropic conductivity model, as illustrated in Fig.~\ref{fig:MultiLayer}.
The total electric field in the $m$-th dielectric layer is expressed as the superposition of forward- and backward-propagating plane waves, given by
\begin{equation}
	\boldsymbol{\mathrm{E}}_m=(E^{+}_{x,m}\mathbf{\hat{x}} + E^{+}_{y,m}\mathbf{\hat{y}})e^{ikz} + (E^{-}_{x,m}\mathbf{\hat{x}} + E^{-}_{y,m}\mathbf{\hat{y}})e^{-ikz},
\end{equation}
where $\left(E^{+}_{x,m},E^{+}_{y,m}\right)$ and $\left(E^{-}_{x,m},E^{-}_{y,m}\right)$ are complex amplitudes of the forward $(+\mathbf{\hat{z}})$ and backward $(-\mathbf{\hat{z}})$ waves, respectively. We adopt the time dependence convention $e^{-i\omega t}$, and fields are described by their phasors. Then, the corresponding total magnetic field is given by

\begin{equation}
	\boldsymbol{\mathrm{H}}_m= \eta_m^{-1} (E^{+}_{x,m}\mathbf{\hat{y}} - E^{+}_{y,m}\mathbf{\hat{x}})e^{ikz} - \eta_m^{-1} (E^{-}_{x,m}\mathbf{\hat{y}} - E^{-}_{y,m}\mathbf{\hat{x}})e^{-ikz},
\end{equation}
where $\eta_m = \eta_0/\sqrt{\varepsilon_{\mathrm{r},m}} = \sqrt{\mu_0/\left(\varepsilon_0\varepsilon_{\mathrm{r},m}\right)}$ is the wave impedance in the $m$-th dielectric layer with relative dielectric constant $\varepsilon_{\mathrm{r},m}$. In our proposed platform, we consider a single 2D material placed at the center of the cavity $(z=0)$, sandwiched between two dielectric layers labeled by the index $m=1,2$, as illustrated in Fig. \ref{fig:MultiLayer}.

By applying the electric and magnetic boundary conditions at the interface between the first and second layers, accounting for the presence of the anisotropic 2D material, and using the state vector in the $m$-th layer defined as $\boldsymbol{\Psi}_m=\left[E^{+}_{x,m}, E^{+}_{y,m}, E^{-}_{x,m}, E^{-}_{y,m} \right]^\mathrm{T}$, the 2D material interaction with the electromagnetic wave is described by 
\begin{equation}
	\boldsymbol{\Psi}_{2} = 
	\overline{\boldsymbol{\mathrm{T}}}_{\sigma}
	\boldsymbol{\Psi}_1,
\end{equation}
where the forward transfer matrix is

\begin{equation}
	\overline{\boldsymbol{\mathrm{T}}}_{\sigma} = \frac{\eta_{2}}{2}
	\left[\begin{array}{cc}
		\left(\eta_{2}^{-1} + \eta_1^{-1}\right)\overline{\mathbf{I}} - \overline{\boldsymbol{\sigma}} & 
		\left(\eta_{2}^{-1} - \eta_1^{-1}\right)\overline{\mathbf{I}} - \overline{\boldsymbol{\sigma}}\\
		\left(\eta_{2}^{-1} - \eta_1^{-1}\right)\overline{\mathbf{I}} + \overline{\boldsymbol{\sigma}} & 
		\left(\eta_{2}^{-1} + \eta_1^{-1}\right)\overline{\mathbf{I}} + \overline{\boldsymbol{\sigma}}\\
	\end{array}\right],
\label{eq:TM_sigma}
\end{equation}
and $\overline{\mathbf{I}}$ is the $2 \times 2$ identity matrix. The transfer matrices for the dielectric layers, $\overline{\boldsymbol{\mathrm{T}}}_{\rm d}$, and for the partial reflective mirrors at both ends, modeled by two DBRs, $\overline{\boldsymbol{\mathrm{T}}}_{\rho}$, corresponding to the structure shown in Fig.~\ref{fig:Structure}, are provided in the Supporting Information. The field reflection coefficient of the partial reflecting mirrors is $\rho = \rho_0e^{i\theta_{\rho}}$, where $\rho_0$ and $\theta_\rho$ represent the reflection magnitude and phase, respectively \cite{furman2023frozen}.
Therefore, for the structure illustrated in Fig. \ref{fig:Structure}, the total transfer matrix is written as

\begin{equation}
    \overline{\boldsymbol{\mathrm{T}}}_{\mathrm{tot}} = 
    \overline{\boldsymbol{\mathrm{T}}}_{\rho}
    \overline{\boldsymbol{\mathrm{T}}}_{\rm d}
    \overline{\boldsymbol{\mathrm{T}}}_{\sigma}
    \overline{\boldsymbol{\mathrm{T}}}_{\rm d}
    \overline{\boldsymbol{\mathrm{T}}}_{\rho}.
\label{eq:total_T}
\end{equation}
From this total transfer matrix, the reflection and transmission matrices are derived, allowing the calculation of reflectance and transmittance for any incident polarization. The total $4\times4$ transfer matrix $\overline{\boldsymbol{\mathrm{T}}}_{\mathrm{tot}}$ is partitioned into $2\times2$ submatrices as

\begin{equation}
  \overline{\boldsymbol{\mathrm{T}}}_\mathrm{tot} = 
	\left[\begin{array}{cc}
			\overline{\boldsymbol{\mathrm{T}}}_{11} & 
			\overline{\boldsymbol{\mathrm{T}}}_{12}\\	
			\overline{\boldsymbol{\mathrm{T}}}_{21} & 
			\overline{\boldsymbol{\mathrm{T}}}_{22}\\
		\end{array}\right].
\end{equation}
These submatrices form a system of equations relating the incident, reflected, and transmitted fields shown in Fig. \ref{fig:Structure}, expressed as

\begin{equation}
\left[\begin{array}{c}
\boldsymbol{\mathrm{E}}_{\mathrm{t}} \\
\boldsymbol{\mathrm{0}}
\end{array}\right]= \left[\begin{array}{cc}
			\overline{\boldsymbol{\mathrm{T}}}_{11} & 
			\overline{\boldsymbol{\mathrm{T}}}_{12}\\	
			\overline{\boldsymbol{\mathrm{T}}}_{21} & 
			\overline{\boldsymbol{\mathrm{T}}}_{22}\\
		\end{array}\right] \left[\begin{array}{c}
\boldsymbol{\mathrm{E}}_{\mathrm{inc}} \\
\boldsymbol{\mathrm{E}}_{\mathrm{r}}
\end{array}\right].
\end{equation}
By definition, the transmitted and reflected fields are related to the incident field as $\boldsymbol{\mathrm{E}}_{\mathrm{t}} = \overline{\boldsymbol{\mathrm{t}}} \: \boldsymbol{\mathrm{E}}_{\mathrm{inc}}$ and $\boldsymbol{\mathrm{E}}_{\mathrm{r}} = \overline{\boldsymbol{\mathrm{\rho}}} \: \boldsymbol{\mathrm{E}}_{\mathrm{inc}}$. Consequently, the reflection and transmission matrices are given by

\begin{gather}
    \overline{\boldsymbol{\mathrm{\rho}}} = -\overline{\boldsymbol{\mathrm{T}}}_{22}^{-1} \: \overline{\boldsymbol{\mathrm{T}}}_{21}, \\
    \overline{\boldsymbol{\mathrm{t}}} = \overline{\boldsymbol{\mathrm{T}}}_{11} - \overline{\boldsymbol{\mathrm{T}}}_{12} \: \overline{\boldsymbol{\mathrm{T}}}_{22}^{-1} \: \overline{\boldsymbol{\mathrm{T}}}_{21}.
\end{gather}
Then, the reflectance and transmittance power for an incident wave $\mathbf{E}_{\mathrm{inc}}$ of arbitrary polarization are given by

\begin{gather}
    \mathcal{T}=\frac{|\mathbf{E}_{\mathrm{t}}|^2}{|\mathbf{E}_{\mathrm{inc}}|^2} = \frac{\mathbf{E}_{\mathrm{inc}}^{\dagger}\: \overline{\boldsymbol{\mathsf{T}}} \: \mathbf{E}_{\mathrm{inc}}}{\mathbf{E}_{\mathrm{inc}}^{\dagger}\: \mathbf{E}_{\mathrm{inc}}}, \\
    \mathcal{R}=\frac{|\mathbf{E}_{\mathrm{r}}|^2}{|\mathbf{E}_{\mathrm{inc}}|^2} = \frac{\mathbf{E}_{\mathrm{inc}}^{\dagger}\: \overline{\boldsymbol{\mathsf{R}}} \: \mathbf{E}_{\mathrm{inc}}}{\mathbf{E}_{\mathrm{inc}}^{\dagger}\: \mathbf{E}_{\mathrm{inc}}}.
\end{gather}
Here, $\overline{\boldsymbol{\mathsf{T}}} = \overline{\boldsymbol{\mathrm{t}}}^{\dagger} \: \overline{\boldsymbol{\mathrm{t}}}$ and $\overline{\boldsymbol{\mathsf{R}}} = \overline{\boldsymbol{\mathrm{\rho}}}^{\dagger}  \: \overline{\boldsymbol{\mathrm{\rho}}}$, where ${\dagger}$ denotes the conjugate transpose. Consequently, absorptance is defined as $\mathcal{A} = 1-\mathcal{R-T}$. Having obtained the transmittance and reflectance of the structure shown in Fig. \ref{fig:Structure}, we can evaluate the occurrence of amplification by analyzing how these quantities change with variations in the 2D material parameters. In particular, a negative absorptance indicates that the optical wave extracts energy from the 2D material, leading to optical amplification and gain.

\subsection{Amplified Transmittance and Negative Absorptance}
\begin{figure*}[t]
    \centering
    \includegraphics[width=0.97\linewidth]{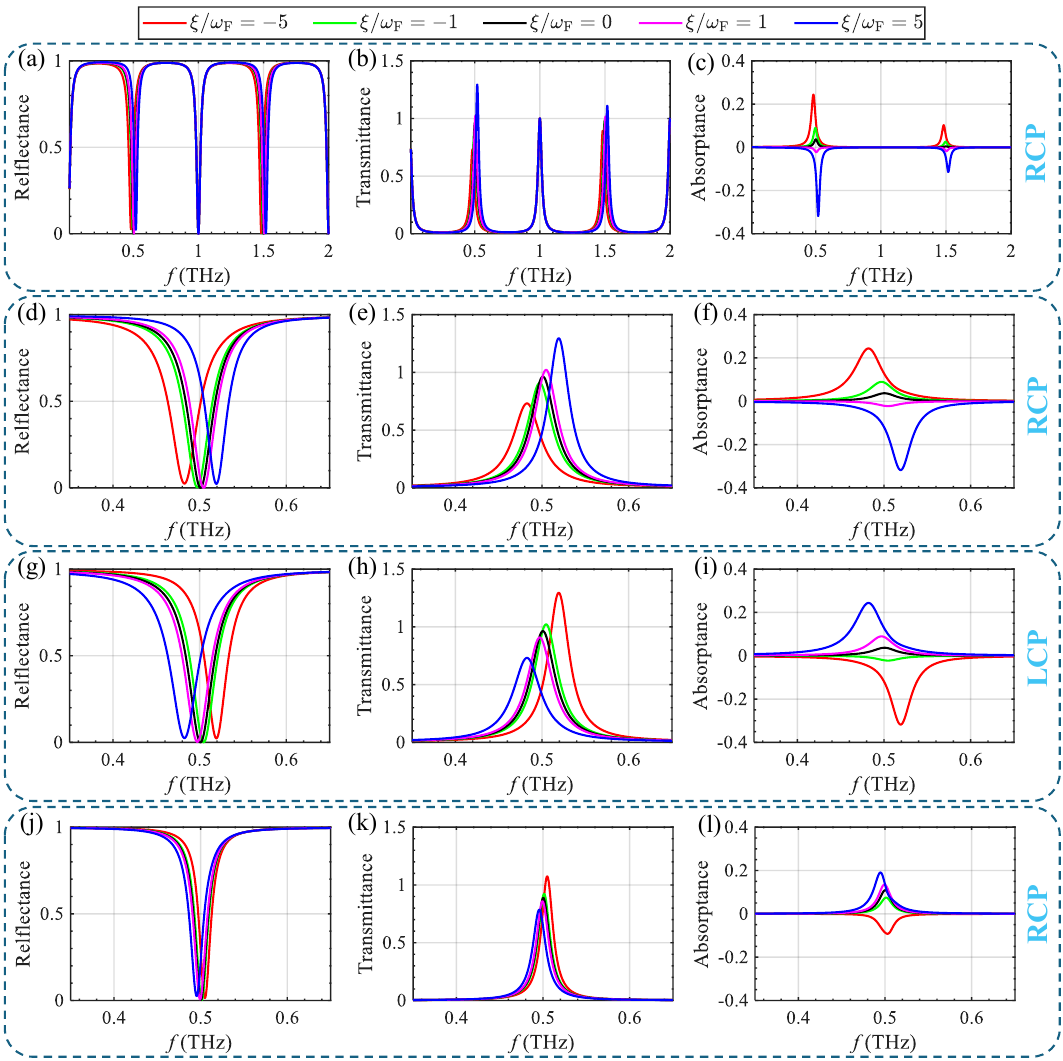}
    \caption{Reflectance, transmittance, and absorptance of the platform proposed in Fig.~\ref{fig:Structure} for various values of the gain parameter $\xi$ of the TBG layer located at the center. DBR mirrors are at both ends as shown in Fig. \ref{fig:Structure}. The first row (a–c) shows results for an RCP incident wave. The second row (d–f) provides a zoomed view near the fundamental resonance at 0.5 THz. The third row (g–i) presents the results for an LCP incident wave. The parameters for (a–i) are: $\varepsilon_{\rm d}=4$, $\omega_{\rm F}/(2\pi)=0.24\:\mathrm{THz}$, $\gamma=10^{12}\:{\rm s}^{-1}$, $\rho=-0.9$, and $L = 149.90 \: \mu{\rm m}$. The fourth row (j–l) corresponds to an RCP incident wave in the same cavity, but with increased TBG losses ($\gamma=5\times10^{12}\:{\rm s}^{-1}$) and higher DBR reflectivity ($\rho=-0.95$). Amplification is seen in various cases.}
    \label{fig:Results_T_R_A}
\end{figure*}

We evaluate the reflectance, transmittance, and absorptance of the structure shown in Fig. \ref{fig:Structure}. Before presenting the results, we first determine an appropriate cavity length. According to the findings in \cite{rappoport2023engineering}, TBG exhibits its maximum amplification around 0.5 THz for $\gamma = 10^{12}\:{\rm s^{-1}}$ and around 1.5 THz for $\gamma = 5 \times 10^{12}\:{\rm s^{-1}}$. Therefore, we choose two FP cavity lengths such that the fundamental resonances correspond to these two frequencies. 
The resonant mode frequencies of an ideal FP cavity of length $L$ (without TBG layer in the middle) are given by $f_q = q v_{\rm p} / (2L)$, where $v_{\rm p} = c/ \sqrt{\varepsilon_{\rm d}}$ is the phase velocity in the dielectric medium inside the cavity. To place the fundamental resonance at $0.5\:\mathrm{THz}$ or $1.5\:\mathrm{THz}$, the required cavity length is $L=c/(2f_1\sqrt{\varepsilon_{\rm d}})$ where $f_1$ denotes the target fundamental resonance frequency of the FP cavity. For $\varepsilon_{\rm d}=4$, this yields $L = 149.90 \:\mu{\rm m}$ for $f_1=0.5 \: {\rm THz}$ and $L = 49.96 \:\mu{\rm m}$ for $f_1=1.5 \: {\rm THz}$.

The TBG parameters used in the analysis are $\omega_{\rm F}/(2\pi) = 0.24\:\mathrm{THz}$, $\gamma = 10^{12}\:{\rm s}^{-1}$, and the field reflection coefficient of the both DBRs is assumed to be $\rho=-0.9$, corresponding to $\rho_0 = 0.9$ with a constant phase $\theta_{\rho} = \pi$ (i.e. 90\% field reflectivity). Note that transmittance values greater than unity or negative absorptance indicate gain, i.e., amplification of the propagating wave.

\begin{figure*}[t]
    \centering
    \includegraphics[width=0.9\linewidth]{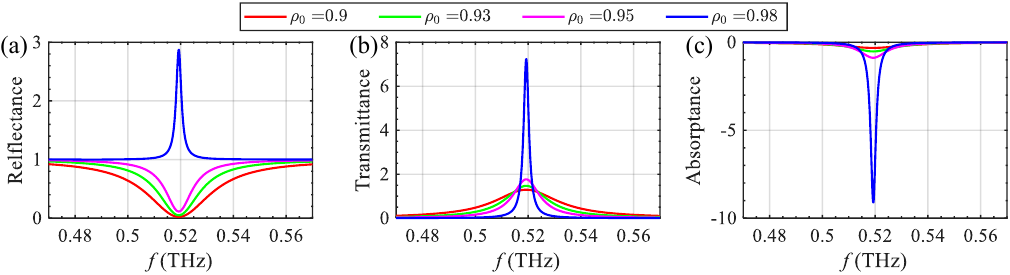}
    \caption{(a) Reflectance, (b) transmittance, and (c) absorptance of the proposed FP cavity when it is illuminated by an RCP incident field. Results are plotted for various values of the DBRs' reflectivity coefficient $\rho_0$. The parameters are: $\varepsilon_{\rm d}=4$, $\omega_{\rm F}/(2\pi)=0.24\:\mathrm{THz}$, $\gamma=10^{12}\:{\rm s}^{-1}$, $\xi/\omega_{\rm F} = 5$, and $L=149.90 \: \mu {\rm m}$.}
    \label{fig:Results_T_R_A_varying_rho0}
\end{figure*}

Fig. \ref{fig:Results_T_R_A}(a-c) (first row) presents the reflectance $\mathcal{R}$, transmittance $\mathcal{T}$, and absorptance $\mathcal{A}$ of the structure shown in Fig. \ref{fig:Structure} for an RCP incident wave from the left as the gain parameter $\xi$ is varied. Throughout this paper, a circularly polarized incident wave traveling along $+\hat{\mathbf{z}}$  is defined as $\mathbf{E}_{\rm inc} = \frac{1}{\sqrt{2}} \left[ 1, \,  \pm i \right]^{\rm T}$, with the plus (minus) sign corresponding to RCP (LCP). It is observed that for all values of $\xi$ there is no noticeable change around 1 THz and 2 THz, which correspond to the even-order resonances of the FP cavity, denoted by $q=2$ and $q=4$, respectively. As expected, the field profile of these modes is antisymmetric, resulting in an electric-field node at the cavity center, i.e., the standing wave does not interact with the 2D material located at the cavity center. As a result, no amplification or attenuation occurs with variations in $\xi$ at those frequencies.

In contrast, around the odd-order resonances of the FP cavity ($q$ odd), located around 0.5 THz ($q=1$) and 1.5 THz ($q=3$), amplification is observed for positive values of $\xi$ and attenuation for negative values. Fig. \ref{fig:Results_T_R_A}(d-f) (second row) provides a closer view of the cavity response to RCP light near 0.5 THz. These plots reveal that increasing $\xi$, \textcolor{black}{which corresponds to increasing the DC bias applied to the 2D material,} enhances the gain and amplification of the transmitted wave. For positive $\xi$, the absorptance becomes negative, indicating that the 2D material provides energy to the light, and the transmittance exceeds unity, confirming amplification.

Figure \ref{fig:Results_T_R_A}(g-i) (third row) presents the same results for an incident wave with LCP. In this case, the results indicate that gain and amplification can be achieved for negative values of the parameter $\xi$. This demonstrates that by changing the sign of $\xi$, which corresponds to reversing the direction of the DC bias in the structure, the handedness of the amplified wave can be selected. This effect results in chiral light amplification, arising from the chiral nature of the gain, which is associated with the BCD in TBG. This feature highlights the tunability of our structure, enabling control over THz wave polarization. Furthermore, Fig. \ref{fig:Results_T_R_A}(j-l) (fourth row) shows the results for \textcolor{black}{RCP incident light} when both losses and DBR reflectivity are increased. Specifically, the scattering rate is increased to $\gamma = 5 \times 10^{12}\:\mathrm{s}^{-1}$ and the reflectivity to $\rho = -0.95$. Larger losses reduce gain and amplification; however, the higher DBR reflectivity partly compensates by strengthening the cavity resonance. Also, increasing $\xi$ to higher values can offset the effect of high losses, restoring amplification. Note that in all panels of Fig. \ref{fig:Results_T_R_A}, the cavity length is fixed at $L=149.90 \: \mu {\rm m}$.

Figure \ref{fig:Results_T_R_A_varying_rho0} shows the response as the magnitude of the DBR field reflectivity $\rho_0$ is varied while keeping the phase fixed at $\pi$. As seen, higher reflectivity significantly increases the transmittance and makes the absorptance more negative, indicating stronger light–matter interaction and an enhanced cavity effect. Interestingly, the reflectance also exhibits a degree of amplification for $\rho_0 = 0.98$. This enhancement further allows operation with larger loss values ($\gamma$) and smaller gain parameters ($\xi$), reducing the required DC bias applied to the 2D material. \textcolor{black}{Note that varying the phase of the DBR field reflectivity, $\theta_\rho$, primarily results in a slight shift of the cavity resonance frequencies, while leaving the overall spectral features shown in Fig. \ref{fig:Results_T_R_A} essentially unchanged. This behavior follows from the standard resonance condition of optical cavities, in which the reflection phase contributes to the round-trip phase \cite{yariv2007photonics}.}

\begin{figure}
    \centering
    \includegraphics[width=1\linewidth]{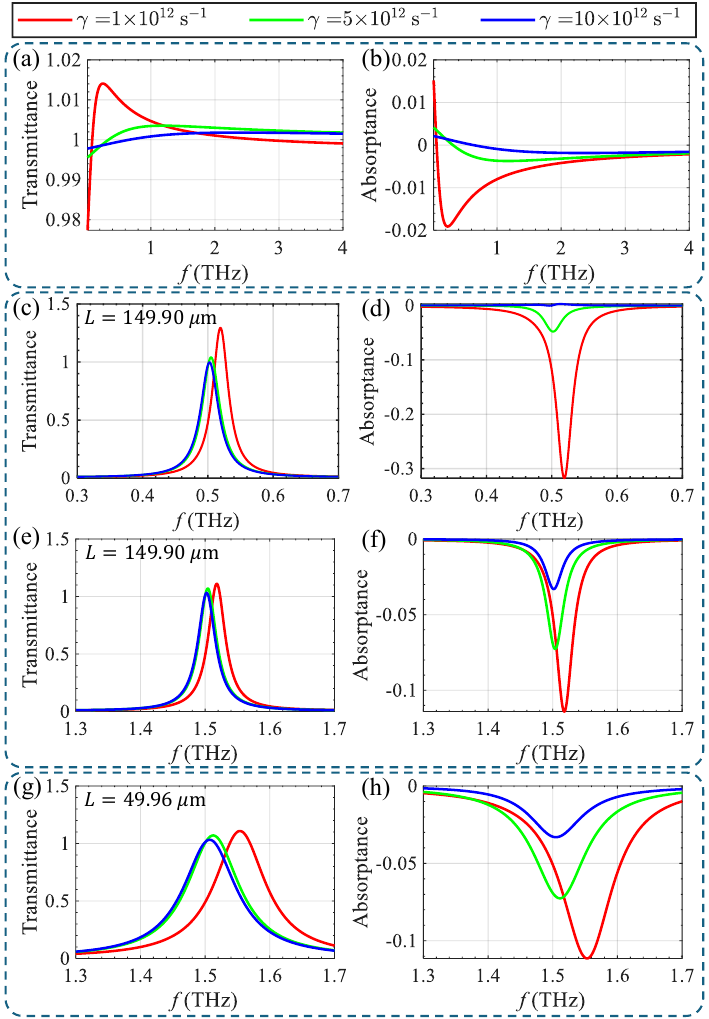}
    \caption{Transmittance (left column) and absorptance (right column) of an electromagnetic structure that includes a TBG layer and is illuminated with an RCP wave. The structure is composed of (a,b) a single TBG in free-space; (c,d) the FP cavity shown in Fig.~\ref{fig:Structure} with $L = 149.90 \: \mu {\rm m}$, corresponding to having the first resonance ($q=1$) near $0.5\:{\rm THz}$; (e,f) the same cavity but observing frequencies near the third resonance ($q=3$); (g,h) the FP cavity with $L = 49.96 \: \mu {\rm m}$, observing frequencies near the fundamental resonance ($q=1$). Other parameters are $\varepsilon_{\rm d}=4$, $\omega_{\rm F}/(2\pi)=0.24\:\mathrm{THz}$, $\xi/\omega_{\rm F} = 5$, and $\rho = -0.9$. Results are plotted for three distinct values of the loss parameter $\gamma$.}
    \label{fig:gamma_effect}
\end{figure}

Next, we examine how variations in the loss parameter $\gamma$ affect the transmittance and gain of both the single TBG (no cavity) and TBG with cavity transmittance response. Figure \ref{fig:gamma_effect} presents the transmittance and absorptance under RCP illumination from the left of the cavity, shown for three distinct loss parameters $\gamma$. Results in the first row (a,b) are for a TBG without a cavity, where the maximum amplification clearly depends on $\gamma$. In contrast, results in the second row (c,d) are for the TBG with cavity designed to have its first ($q=1$) resonance near $0.5\:{\rm THz}$, corresponding to $L = 149.90 \: \mu {\rm m}$. Unlike the single TBG case, the cavity resonance dictates the transmittance-peak frequency, so variations in $\gamma$ do not significantly shift the frequency of maximum transmittance. As expected, an increased $\gamma$ reduces the overall amplification.
The third row (e,f) corresponds to the same cavity operating near the third resonance ($q=3$). For small $\gamma$ (red curve), the transmittance decreases relative to the fundamental resonance, consistent with the single-TBG behavior in the first row. However, when the loss is large (blue curve), the amplification is stronger at this third resonance than at the fundamental resonance.

The fourth row (g,h) shows results for a cavity designed to have the fundamental resonance near $1.5\:{\rm THz}$, corresponding to $L = 49.96 \: \mu {\rm m}$. These results mirror those of the third row but are spectrally expanded, illustrating the scalability of the design with cavity length. Overall, variations in $\gamma$, cavity length, and mode index confirm that the proposed cavity exhibits robust and scalable performance.

\begin{figure}
    \centering
    \includegraphics[width=1\linewidth]{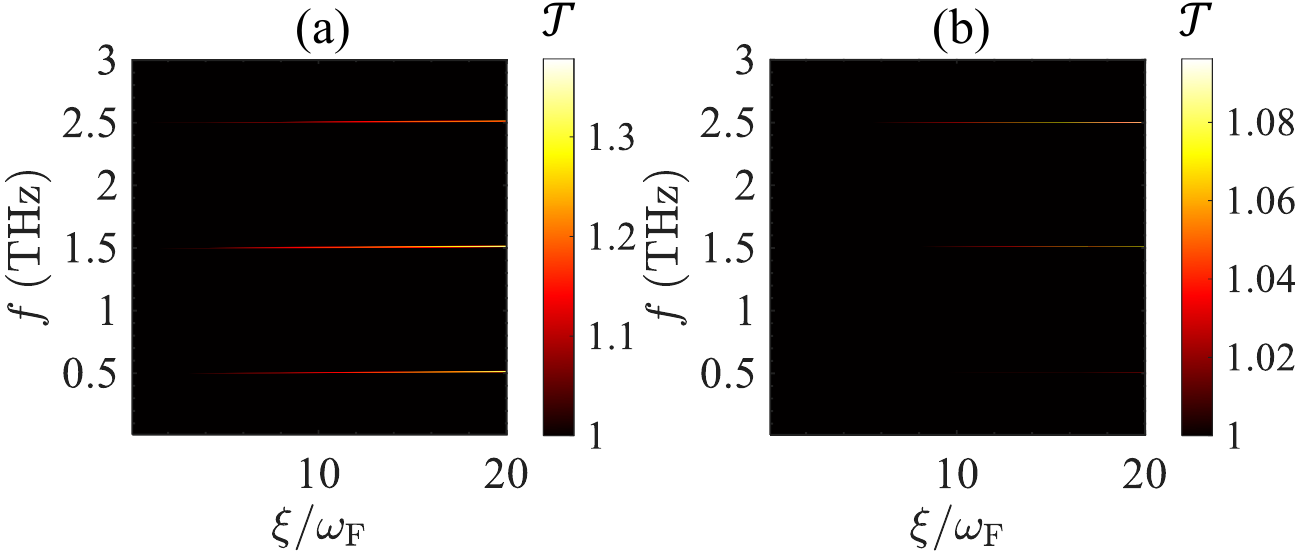}
    \caption{Transmittance under RCP incidence by varying frequency and gain parameter $\xi/\omega_{\rm F}$ when (a) $\gamma = 5\times10^{12} \: {\rm s^{-1}}$, (b) $\gamma = 20 \times 10^{12} \: {\rm s^{-1}}$. Other parameters are $\varepsilon_{\rm d}=4$, $\omega_{\rm F}/(2\pi)=0.24\:\mathrm{THz}$, $\rho = -0.9$, and $L = 149.90 \: \mu {\rm m}$. For large losses $\gamma$, transmittance can still be larger than unity for resonances around $1.5$ THz and $2.5$ THz for a moderate value of the gain parameter.}
    \label{fig:T_varying_xi}
\end{figure}

To better understand the role of losses and to demonstrate that the cavity can still exhibit amplified transmittance at high scattering rates, we plot the transmittance as a function of frequency and gain parameter $\xi$ for larger loss values in Fig.~\ref{fig:T_varying_xi}.
Figure~\ref{fig:T_varying_xi}(a) shows that, for a moderate loss value of $\gamma = 5 \times 10^{12}\,\mathrm{s^{-1}}$, the amplification increases with the gain parameter around all odd-order resonances. However, the onset of amplified transmittance occurs at smaller $\xi$ values around $1.5\: {\rm THz}$ compared to the other resonances.
In contrast, Fig.~\ref{fig:T_varying_xi}(b), corresponding to a higher loss value of $\gamma = 20 \times 10^{12},\mathrm{s^{-1}}$, shows that amplification occurs primarily around 1.5 and 2.5~THz, while amplification near 0.5~THz is observed only for very large values of the gain parameter. Moreover, the onset of amplification appears at a smaller $\xi$-threshold value for 2.5~THz, which corresponds to the higher odd-order longitudinal cavity mode ($q=5$).

The numerical observations relative to the gain parameter $\xi$, frequency $f$, and TBG loss $\gamma$ are consistent with the analytical prediction in Eq.~\eqref{eq:gain_condition}, which determines the required $\xi$ value for the onset of amplification in TBG. In the regime $\omega \gg \gamma$, increasing $\gamma$ raises the required $\xi$ to initiate amplification, whereas in the regime $\omega \ll \gamma$, the required $\xi$ decreases with increasing frequency. Consequently, higher-order longitudinal resonance modes begin amplifying at lower $\xi$ values.
These results indicate that, in general, as the loss parameter increases, higher odd-order longitudinal cavity modes are required to achieve the onset of amplification when tuning the gain parameter $\xi$.

The Supporting Information further validates the scalability and robustness of the cavity-TBG through a systematic optimization study, where the transmittance is mapped as a function of both cavity length and incident frequency for different loss values $\gamma$. In addition, the polarization of the transmitted wave is examined by evaluating the co- and cross-polarized field components versus frequency, gain, and loss parameters for both RCP and LCP incident waves. The cross-polarized component is found to be negligible, indicating that the cavity preserves the polarization of the incident wave. This means that the main function of cavity-TBG is amplification rather than polarization conversion, because the input polarization is maintained alongside scalable amplification.

In summary, analyzing the reflectance, transmittance, and absorptance of the structure shown in Fig. \ref{fig:Structure} reveals that, for an RCP incident wave from the left, the cavity can produce an amplified transmitted wave, with the level of amplification controlled by the parameter $\xi$, which is related to the applied DC bias on the 2D material. Furthermore, by reversing the sign of the applied bias, one can switch the amplification between the RCP and LCP incident waves. Also, if an RCP illuminates the cavity with TBG at its center from the right side, amplification for positive values of $\xi$ would occur for LCP because of the chiral properties of TBG. 
Moreover, inherent losses in the 2D material reduce amplification, but this can be compensated by increasing the reflectivity of the DBRs in the cavity, increasing the $\xi$, or operating at higher odd-order longitudinal cavity modes. Finally, it is worth noting that any low-symmetry 2D material exhibiting a BCD-induced gain could replace TBG in this configuration.

\section{Lasing Cavity Threshold}
\label{sec:lasing}

The reflectance, transmittance, and absorptance analysis presented in Section \ref{sec:T_and_R} showed that the 2D material with BCD can provide optical gain through its interaction with the intracavity field. \textcolor{black}{Following the results shown in Fig. \ref{fig:Results_T_R_A}, when $\mathcal{T}>1$, the gain provided by the 2D material leads to amplification, resulting in an amplified transmitted wave. This amplified transmittance may lead to instability of the cavity. Such instability occurs when the wave undergoing a round trip inside the cavity has gain that overcompensates for the cavity's total losses, corresponding to the onset of self-sustained oscillation and lasing.} To this purpose, we perform a modal analysis to determine the eigenmodes of the cavity loaded with the 2D material at its center, as illustrated in Fig. \ref{fig:Cavity_Dispersion}. \textcolor{black}{The cavity considered in Fig. \ref{fig:Cavity_Dispersion} is a slightly modified version of the structure shown in Fig. \ref{fig:Structure}, in which the left DBR is replaced by a perfect mirror, whereas the DBR on the right provides field reflectivity $\rho$. This modification eliminates the frequency dispersion associated with the DBRs without compromising the physics considerations pertaining to lasing, thereby avoiding the need to design the DBRs specifically for the lasing frequency and simplifying the analysis.}
The complex-frequency resonance condition of the cavity is then determined, and the lasing threshold is identified from the vanishing of the imaginary part of the complex frequency.

\subsection{Complex-Frequency Resonance Condition}

\begin{figure}
    \centering
    \includegraphics[width=1\linewidth]{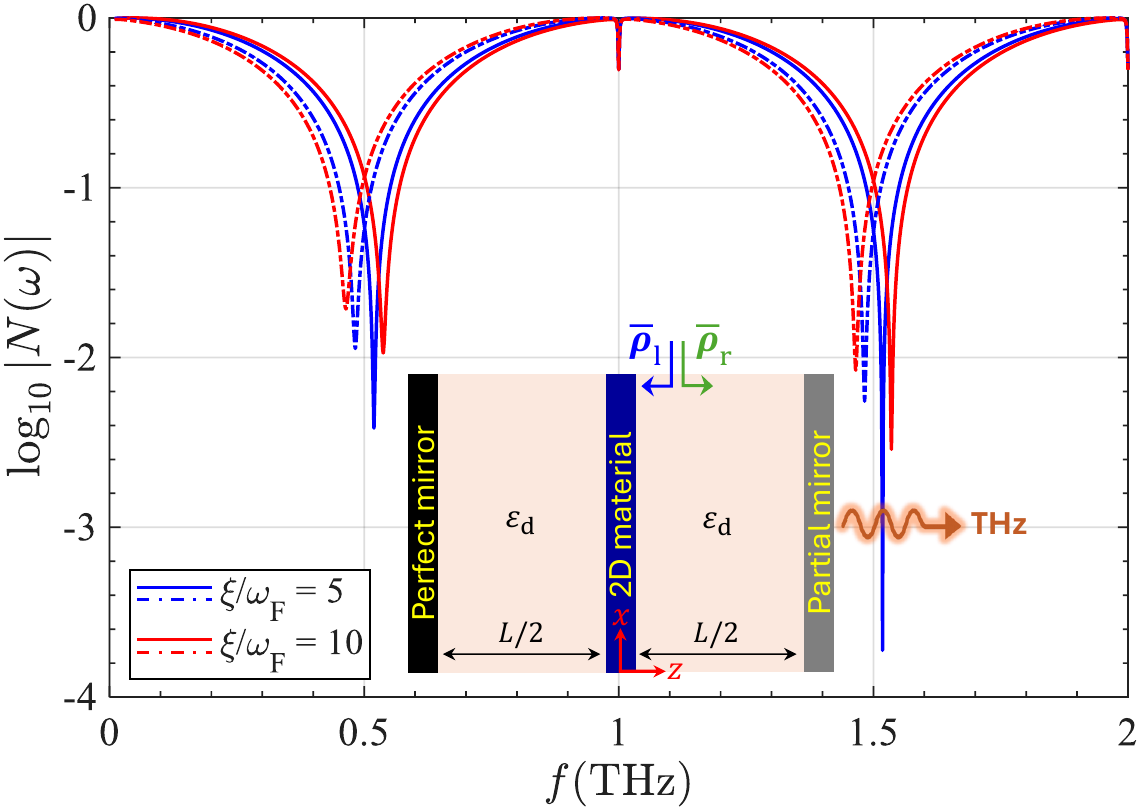}
    \caption{Real-valued resonance frequencies given by the dips of $|N(\omega)|$ for two distinct values of $\xi$. The solid curves correspond to the minus-sign branch, whereas the dash-dotted curves correspond to the plus-sign branch of the resonance condition $N(\omega)=0$ in Eq. \eqref{eq:modes}. To fully satisfy $N(\omega)=0$, $f$ must be complex \textcolor{black}{(except when $\xi=\xi_{\rm th}$)}; however, the dips for real $f$ in this plot approximate the cavity's oscillation frequencies. Parameters used are $\varepsilon_{\rm d}=4$, $\omega_{\rm F}/(2\pi)=0.24\:\mathrm{THz}$, $\gamma=10^{12}\:\mathrm{s}^{-1}$, $L = 149.90 \: \mu {\rm m}$, and $\rho=-0.99$, representing the field reflectivity of the partial mirror on the right. The inset depicts the proposed cavity configuration, as in Fig.~\ref{fig:Structure}, but terminated by a perfect (fully reflective) mirror on the left and a partially reflective mirror on the right.}
    \label{fig:Cavity_Dispersion}
\end{figure}

Let $\mathbf{E}$ denote the electric field at an arbitrary position $z$ inside the cavity. The resonance condition is given by

\begin{equation}
    \overline{\boldsymbol{\rho}}_{\mathrm{r}} \overline{\boldsymbol{\rho}}_{\mathrm{l}} \mathbf{E} = \mathbf{E},
    \label{eq:resonance}
\end{equation}
where $\overline{\boldsymbol{\rho}}_{\mathrm{r}}$ and $\overline{\boldsymbol{\rho}}_{\mathrm{l}}$ are the reflection matrices looking to the right and to the left, respectively, evaluated at a chosen arbitrary $z$ location. Here, we take this location to be immediately to the right of the 2D material, as illustrated in Fig. \ref{fig:Cavity_Dispersion}. Consequently, the resonance condition leads to

\begin{equation}
    F(\omega) \equiv \det \left(\overline{\boldsymbol{\rho}}_{\mathrm{r}} \overline{\boldsymbol{\rho}}_{\mathrm{l}} - \overline{\textbf{I}} \right) = 0.
\label{eq:det}
\end{equation}
The determination of $\overline{\boldsymbol{\rho}}_{\mathrm{r}}$ and $\overline{\boldsymbol{\rho}}_{\mathrm{l}}$ is detailed in the Supporting Information. Based on this analysis, the cavity resonances \textcolor{black}{associated with the zeros of the numerator of $F(\omega)$}, are described by the following condition:

\begin{equation}
        \textcolor{black}{N(\omega)} \equiv \sigma_{1,2}  +i \frac{1}{\eta_{\mathrm{d}}} \frac{1 -  \cot^2(k_{\mathrm{d}} L/2)  +2i y_{\mathrm{m}} \cot(k_{\mathrm{d}} L/2) }{iy_{\mathrm{m}} - \cot(k_{\mathrm{d}} L/2)} = 0,
\label{eq:modes}
\end{equation}
where $\eta_{\mathrm{d}} = \eta_0/\sqrt{\varepsilon_{\mathrm{d}}}$ is the characteristic wave impedance of the dielectric, and $y_{\mathrm{m}} = (1-\rho)/(1+\rho)$ characterizes the reflectivity $\rho$ of the partial mirror on the right side of the cavity, as illustrated in Fig. \ref{fig:Cavity_Dispersion}. In the limit $\rho \to -1$, the mirror becomes perfectly reflecting and $y_{\mathrm{m}}$ diverges to infinity. In addition to the modes \textcolor{black}{with a maximum of electric field at the center of the cavity} described by Eq. \eqref{eq:modes}, the modal analysis yields another set of complex-frequency cavity resonances associated with the poles arising from the denominator of $F(\omega)$, corresponding to resonances with an antisymmetric field profile, i.e., with a field node at the center of the cavity, as discussed in Section \ref{sec:T_and_R}, where the 2D material is placed. As a result, they experience neither amplification nor attenuation, since the field vanishes at the 2D layer and light–matter interactions are negligible.

We focus on the set of resonances, given by $N(\omega) = 0$ in Eq. \eqref{eq:modes}. As discussed in Section \ref{sec:T_and_R}, these resonances exhibit an electric field maximum at the cavity center, thereby maximizing light-matter interaction. The resonance condition $N(\omega) = 0$ for these resonances yields two distinct branches associated with the minus and plus signs. As proved in the Supporting Information, the resonances in Eq. \eqref{eq:modes} correspond to the polarization eigenstates of the 2D material, given in Eq. \eqref{eq:eigenvectors}. Specifically, the minus-sign branch corresponds to the 2D material eigenstate $\mathbf{E}^{\sigma}_{1}$, while the plus-sign branch corresponds to the eigenstate $\mathbf{E}^{\sigma}_{2}$.

Similar to the modal analysis presented in \cite{hakimi2024chiral}, where a structure incorporating a stack of 2D material yielded two types of eigenmodes, one exhibiting optical gain and the other loss, here we also observe two different branches. One branch experiences optical gain, whereas the other experiences loss. The first two terms of $N(\omega)$ correspond to the eigenvalues of the conductivity matrix given in Eq. \eqref{eq:eigenval}, corresponding to the two polarization eigenstates in  Eq.~(\ref{eq:eigenvectors}). Clearly, the minus-sign branch (Mode 1) can produce a negative real part of the eigenvalue (when $\xi>0$), which may lead to optical gain capable of compensating cavity losses and enabling lasing. Consequently, for $\xi > 0$, we assume that the polarization state of the lasing cavity is $\mathbf{E}^{\sigma}_{1}$. \textcolor{black}{Recalling from Eq.~\eqref{eq:power} that the threshold for amplification is ${\rm Re} \left(\sigma_1 \right)=0$, we observe from Eq.~(\ref{eq:modes}) that this condition alone is not sufficient to establish lasing that requires ${\rm Re}\left( N(\omega) \right)=0$, due to the cavity losses accounted for by the real part of $y_{\mathrm{m}}$. When power is not allowed to exit the cavity and a lossless mirror is used, $y_{\mathrm{m}}$ is purely imaginary, and the threshold of amplification corresponds to the threshold of lasing.}

The dips in Figure \ref{fig:Cavity_Dispersion} show the real-valued resonance frequencies obtained by plotting $\left|N(\omega)\right|$ versus real $f$, considering two distinct values of the gain parameters $\xi$. The solid lines correspond to the minus-sign branch, whereas the dash-dotted lines correspond to the plus-sign branch of the resonance condition in Eq. \eqref{eq:modes}. In particular, the dips correspond approximately to the real part of the complex-valued resonant frequencies of the cavity that fully satisfy $N(\omega)=0$. Interaction between the optical wave propagating inside the cavity and the 2D material slightly shifts the resonance frequencies from the FP cavity resonances near 0.5 THz and 1.5 THz, due to the polarization eigenstate in Eq. \eqref{eq:eigenvectors} involved in the process. Near each of these frequencies, two distinct dips appear, corresponding to the minus-sign (polarization eigenstate 1) and plus-sign (polarization eigenstate 2) branches. The parameters used in this analysis are $\varepsilon_{\rm d}=4$, $\omega_{\rm F}/(2\pi)=0.24\:\mathrm{THz}$, $\gamma = 10^{12}\:{\rm s}^{-1}$, $L = 149.90 \: \mu {\rm m}$, and a field reflection coefficient $\rho=-0.99$ for the partial mirror on the right side of the cavity shown in Fig. \ref{fig:Cavity_Dispersion}, corresponding to 99\% reflectivity.

Having identified the frequencies of the cavity resonances interacting with the 2D material, we expect that Mode 1 (the minus-sign branch), which occurs at frequencies slightly above 0.5 THz and 1.5 THz, may exhibit amplification. In the next section, we verify this behavior by performing a complex frequency analysis. There, we also determine the lasing threshold---that is, the minimum value of the parameter $\xi$ (corresponding to the applied bias) required to initiate lasing.

\subsection{Lasing Threshold}
The resonance frequencies obtained from Eq. \eqref{eq:modes} and shown in Fig. \ref{fig:Cavity_Dispersion} for various values of the $\xi$ parameter are generally complex valued because they account for loss or gain in the system. Using the time-harmonic convention $e^{-i\omega t}$, a positive imaginary part of the complex frequency (${\rm Im}(f) > 0$) indicates an unstable, exponentially growing field, indicating optical amplification. The lasing threshold is found by setting ${\rm Im}(f)=0$, marking the transition from stability to instability.
 
\begin{figure}
    \centering
    \includegraphics[width=0.95\linewidth]{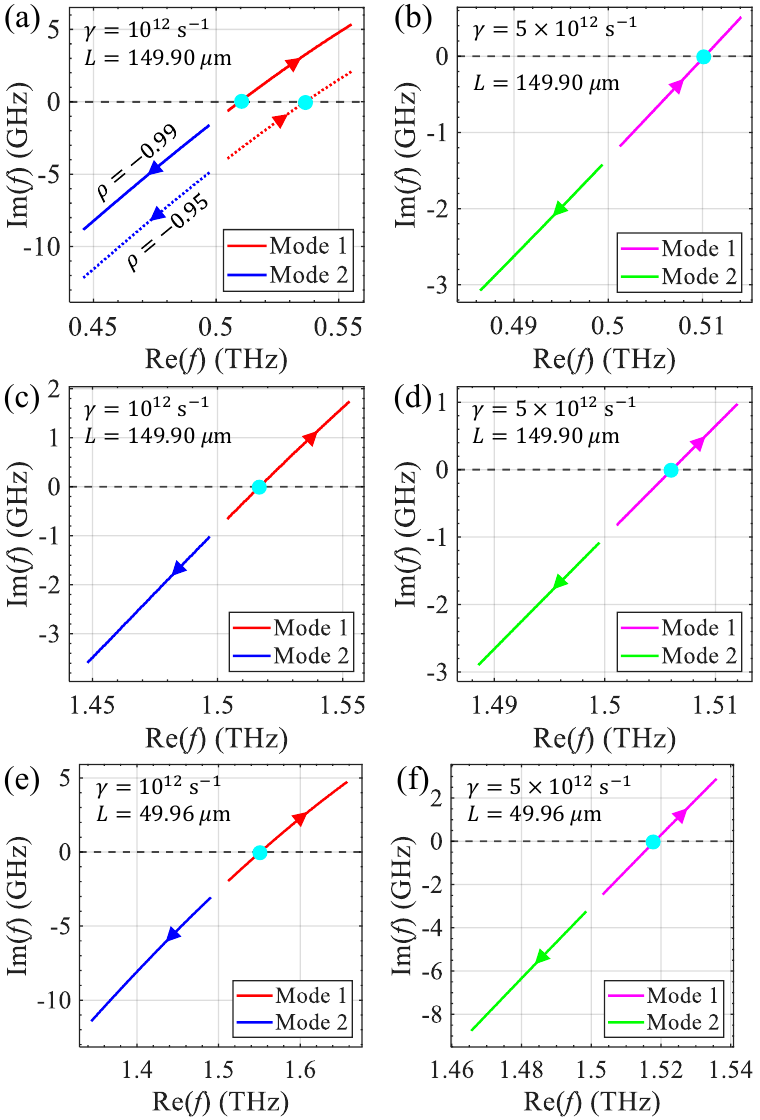}
    \caption{Imaginary part versus real part of the lasing natural frequency as $\xi/\omega_{\rm {\rm F}}$ varies within $1<\xi/\omega_{\rm F}<15$ [Solution to Eq. \eqref{eq:modes}]. The top row shows the lasing around the fundamental resonance ($q=1$), whereas the middle row shows lasing around the third resonance ($q=3$), both for $L = 149.90 \: \mu {\rm m}$. The bottom row shows lasing around the fundamental resonance ($q=1$) for $L = 49.96 \: \mu {\rm m}$. Panels (a), (c), and (e) correspond to $\gamma = 10^{12}$, while (b), (d), and (f) correspond to $\gamma = 5\times 10^{12}$. \textcolor{black}{Solid lines correspond to $\rho=-0.99$, while the dotted lines correspond to $\rho=-0.95$. Other parameters are $\varepsilon_{\rm d}=4$, $\omega_{\rm F}/(2\pi)=0.24\:\mathrm{THz}$.} \textcolor{black}{Instability occurs when ${\rm Im}(f)>0$, with the lasing threshold obtained by ${\rm Im}(f)=0$,} indicated by the intersections of the curves with the dashed line, marked by cyan dots.}
    \label{fig:Complex_frequency}
\end{figure}

Figure \ref{fig:Complex_frequency} shows the complex frequencies of the cavity resonances as the parameter $\xi$ is varied over the range $1<\xi/\omega_{\rm F}<15$. The top row presents the results near the fundamental resonance (around 0.5 THz, for $q=1$) for two distinct values of loss $\gamma$, whereas the middle row corresponds to the second resonance (around 1.5 THz, for $q=3$), both for $L = 149.90 \: \mu {\rm m}$. \textcolor{black}{In panel (a), the reflectivity of the partial reflective mirror is varied, with the solid lines corresponding to $\rho=-0.99$ and the dotted lines corresponding to $\rho=-0.95$.} The bottom row shows results around the fundamental resonance (around 1.5 THz, for $q=1$) when $L = 49.96 \,\mu{\rm m}$. In all panels, as expected, Mode 1, corresponding to the minus branch of Eq. \eqref{eq:modes} and the polarization state $\mathbf{E}^{\sigma}_{1}$ in Eq. \eqref{eq:eigenvectors}, exhibits instability (i.e., amplification) as $\xi$ increases. 
The onset of lasing, indicated by the points where the curves intersect the ${\rm Im}(f)=0$ dashed line, is marked with cyan dots. These intersection points occur at $\xi_{\rm th}/\omega_{\rm F} = 2.41$ for $\left(\rho=-0.99\right)$ and $9.72$ for $\left(\rho=-0.95\right)$ in (a), $10.78$ in (b), $4.73$ in (c), $7.40$ in (d), $4.80$ in (e), and $7.40$ in (f), representing the minimum gain parameter required to initiate lasing for the given cavity lengths. Table \ref{tab:xi_th} summarizes the corresponding lasing frequencies and gain thresholds for two $\gamma$ values, two cavity lengths, and two partial mirror reflectivity values. \textcolor{black}{For a direct comparison with the multilayer structure including various TBG layers separated by dielectric spacers studied in Ref. \cite{hakimi2024chiral}, a comparable threshold based on the specifications in the first row of Table \ref{tab:xi_th} with $\rho_0=0.99$ would be achieved with a cavity with 22 TBG separated by $5~\mu{\rm m}$ dielectric spacers. In contrast, the present cavity structure achieves lasing using a single TBG, enabled by the enhanced light-matter interaction, thereby substantially reducing the structural and fabrication complexity.}
Moreover, the required gain parameters for the middle and bottom two rows are nearly identical, despite the different cavity lengths and longitudinal resonance indices, $q=3$ and $q=1$, respectively. This behavior is consistent with the results in Fig. \ref{fig:gamma_effect} (third and fourth rows), demonstrating the scalability and tunability of the proposed TBG-loaded cavity concept, where the lasing frequency, cavity length, and oscillating resonance-mode index $q$ can be readily adjusted as desired. For large $\gamma$, the lasing threshold at the second resonance (near 1.5 THz) is lower than at the first resonance (near 0.5 THz).

Furthermore, polarization eigenstate 1 in Eq. \eqref{eq:eigenvectors} is responsible for amplification, consistent with Section \ref{sec:T_and_R}, which predicts amplification for RCP when propagating along $+\hat{\mathbf{z}}$ and for LCP when propagating along $-\hat{\mathbf{z}}$ for $\xi>0$. As a result, the intracavity eigenmode associated with lasing is the polarization eigenstate 1, and consequently, the emitted radiation, coming out from the right of the cavity shown in Fig. \ref{fig:Cavity_Dispersion}, is predominantly RCP. Notably, analogous to Section \ref{sec:T_and_R}, where changing the sign of $\xi$ controls which handedness of the incident wave experiences amplification, here the sign of the applied DC bias can similarly control the handedness of the wave emitted from the lasing cavity. This feature enables the realization of a chiral THz lasing device.

\begin{table}[t]
    \centering
    \caption{Minimum $\xi$ required to initiate lasing and the corresponding lasing frequency for various loss values ($\gamma$), cavity lengths ($L$), \textcolor{black}{and partial-mirror field-reflectivity magnitudes ($\rho_0$)}. The superscript in approximated threshold values is computed using: \eqref{eq:xi_th_approx_high} or \eqref{eq:xi_th_approx_low}. \textcolor{black}{The last column shows the corresponding required DC bias, calculated from $\xi_{\rm th}$ when the TBG is placed as a 2D material with a BCD of $D_{\rm B}=10~\mathrm{nm}$.}}
    \begin{tabular}{|c|c|c|c|c|c|c|}
        \hline
        $L\: (\mu {\rm m})$ & $\gamma \: ({\rm s}^{-1})$ & \textcolor{black}{$\rho_0$} & $f_{\rm r} \: ({\rm THz})$ & $\xi_{\rm th}/\omega_{\rm F}$ & $\xi_{\rm th,a}/\omega_{\rm F}$ & \textcolor{black}{$E_0 \: ({\rm V/m m})$} \\ 
        \hline
        \multirow{5}{*}{149.90} & \multirow{2}{*}{$10^{12}$} & 0.99 & 0.510 & 2.41 & 2.41$^{\eqref{eq:xi_th_approx_high}}$ & 76.84 \\ 
        \cline{3-7}
         &  & 0.95 & 0.536 & 9.72 & 9.47$^{\eqref{eq:xi_th_approx_high}}$ & 301.95 \\ 
        \cline{2-7}
         & $5 \times 10^{12}$ & \multirow{5}{*}{0.99} & 0.510 & 10.78 & 11.39$^{\eqref{eq:xi_th_approx_low}}$ & 343.71 \\ 
        \cline{2-2} \cline{4-7}
        \multirow{2}{*}{} & $10^{12}$ &  & 1.517 & 4.73 & 4.61$^{\eqref{eq:xi_th_approx_high}}$ & 150.81 \\ 
        \cline{2-2} \cline{4-7}
         & $5 \times 10^{12}$ &  & 1.506 & 7.40 & 9.36$^{\eqref{eq:xi_th_approx_high}}$ & 235.95  \\ 
        \cline{1-2} \cline{4-7}
        \multirow{2}{*}{49.96} & $10^{12}$  &  & 1.551 & 4.80 & 4.61$^{\eqref{eq:xi_th_approx_high}}$ & 153.05  \\ 
        \cline{2-2} \cline{4-7}
         & $5 \times 10^{12}$ &  & 1.518 & 7.40 & 9.36$^{\eqref{eq:xi_th_approx_high}}$ & 235.95 \\ 
        \hline
    \end{tabular}
    \label{tab:xi_th}
\end{table}

We estimate the lasing threshold $\xi_{\rm th}$ analytically by applying suitable approximations to Eq. \eqref{eq:modes}. First, since the reflectivity of the cavity’s right partial mirror is relatively high, we have $\left|y_{\rm m} \right| \gg 1$. Additionally, the interaction between the cavity field and the 2D material slightly shifts the lasing frequency from the resonance of the unloaded FP cavity at 0.5 THz ($q=1$) and 1.5 THz ($q=3$). Assuming this shift to be small, we have \textcolor{black}{$k_{\mathrm{d}} L/2 \approx q\pi/2$ with $q=1,3,5, \dots$}, implying $\left|\cot(k_{\mathrm{d}} L/2) \right| \ll 1$. 
The approximated threshold $\xi_{\rm th, a}$ is found under two additional assumptions, $\omega \gg \gamma$ and $\omega \ll \gamma$, leading to, respectively:

\begin{subequations}
\begin{equation}
\begin{split}
    \xi_{\rm th,a} ({\omega \gg \gamma}) \approx 2\left(\frac{\omega_{\rm r,a}}{y_{\mathrm{m}}\eta_{\mathrm{d}}\sigma_0} + \gamma\frac{\omega_{\rm F}}{\omega_{\rm r,a}}\left(1 - \left(\frac{\gamma}{\omega_{\rm r,a}}\right)^2\right)\right) \\ \left(1 - \frac{13}{8}\left(\frac{\gamma}{\omega_{\rm r,a}}\right)^2\right)^{-1},    
\end{split}
\label{eq:xi_th_approx_high}
\end{equation}
\begin{equation}
\begin{split}
    \xi_{\rm th,a} ({\omega \ll \gamma}) \approx 2\sqrt{2} \frac{\gamma}{\omega_{\rm r,a}}\left(\frac{\gamma}{y_{\mathrm{m}}\eta_{\mathrm{d}}\sigma_0} + \omega_{\rm F}\left(1 - \left(\frac{\omega_{\rm r,a}}{\gamma}\right)^2\right)\right) \\ \left(1 - \frac{25}{32}\left(\frac{\omega_{\rm r,a}}{\gamma}\right)^2\right)^{-1}.
\end{split}
\label{eq:xi_th_approx_low}
\end{equation}
\label{eq:xi_th_approx}
\end{subequations}
Here, \textcolor{black}{$\omega_{\rm r, a} = 2\pi q f_1$ with $q=1,3,5,\dots$} is the approximate resonance angular frequency, assumed equal to that of the TBG-unloaded FP cavity. The full derivation of these expressions is provided in the Supporting Information.
The penultimate column of Table \ref{tab:xi_th} lists the $\xi_{\rm th, a}$ values obtained using the approximate formulas in Eq. \eqref{eq:xi_th_approx}, based on the corresponding resonance frequency and loss values in each row. Comparison between the third column, which shows the exact gain parameter values, and the fourth column, which shows the approximate values, indicates very good agreement. The differences arise when the resonance angular frequency $\omega$ and loss $\gamma$ are comparable, violating the assumption that one is much larger than the other.
The approximate formulas in Eq. \eqref{eq:xi_th_approx} provide a practical tool for estimating the required gain parameter, and consequently the necessary applied DC bias to the TBG, for the onset of lasing and optical amplification.

\textcolor{black}{After determining the $\xi_{\rm th}$ required for the onset of the instability leading to lasing and the corresponding resonance frequency, as well as the approximation $\xi_{\rm th,a}$ from Eq. \eqref{eq:xi_th_approx}, which can be used to identify a suitable design value, the required DC electric field applied to the 2D material can be obtained. As discussed earlier, the relation between $\xi$ and the required DC bias $E_0$ is given by $E_0 = \hbar \xi/ \left( \pi e D_{\rm B}\right)$. For a given $\xi$, the only material-dependent parameter in this expression is the BCD, $D_{\rm B}$. Here, we consider TBG with $D_{\rm B}=10\,{\rm nm}$ \cite{rappoport2023engineering}. Using this value, the corresponding required $E_0$ is reported in the last column of Table \ref{tab:xi_th}. Importantly, the required DC bias leading to lasing can be reduced by using materials with larger BCD values \cite{zhang2022giant}. The BCD can potentially be enhanced through several approaches, including Fermi-level tuning \cite{kim2024gate}, strain engineering \cite{qin2021strain}, and symmetry breaking \cite{zhao2023berry}. Twisted bilayer structures also provide a promising platform for enhancing the BCD; for example, a BCD as large as $D_{\rm B}=140\,{\rm nm}$ has been reported in twisted bilayer WTe$_2$ \cite{he2021giant}.}

\textcolor{black}{Finally, for a stable lasing cavity, in addition to the gain and instability mechanism investigated here to initiate lasing, the gain-saturation mechanism should be investigated. As the signal grows due to the gain provided by the active material, the gain must eventually decrease and saturate beyond a certain field amplitude, leading to a stable self-oscillatory coherent emission. Investigating this regime requires modeling the material response when the optical field is no longer weak, where calculating the current using a perturbative solution of the Boltzmann transport equation is no longer valid. For example, in \cite{sur2024fully}, the authors investigate the fully nonequilibrium Hall response in the regime of strong applied electric fields, where the displacement of the Fermi surface from its equilibrium configuration becomes sufficiently large that perturbative approaches break down. In this regime, they show that the generated current no longer increases linearly with the applied electric field and instead begins to saturate. Such a nonlinear saturation of the generated current could provide the necessary mechanism for stabilizing the oscillation and enabling a stable self-oscillatory regime in our proposed cavity.}


\section{Conclusion}
We have proposed and theoretically analyzed a compact cavity architecture that enables direct terahertz amplification and lasing by DC electrical pumping of a low-symmetry 2D material with BCD. Embedding a biased 2D material in an FP cavity enhances light–matter interactions, enabling efficient conversion of DC bias into coherent THz radiation. Unlike conventional gain media, this approach does not rely on population inversion and instead exploits intraband Bloch-electron dynamics driven by BCD-induced chiral gain.

Using a transfer-matrix formalism and nonreciprocal modal analysis, we demonstrated selective amplification at the odd-order cavity resonances, negative absorptance, and transmittance exceeding unity, with gain controlled by the bias-dependent parameter $\xi$. We further showed that reversing the DC bias switches the handedness of the amplified or emitted radiation, enabling electrically tunable chiral THz operation. We also investigated the effect of 2D material losses and showed that, even under large intrinsic dissipation, the cavity and gain parameter can compensate for these losses and enable net THz amplification.
Complex-frequency analysis revealed clear lasing thresholds for realistic material losses and mirror reflectivities. We derived approximate analytical expressions that accurately predict the gain threshold. Importantly, our results show that a single 2D material suffices to achieve substantial amplification and lasing, significantly simplifying the structure compared to multilayer designs previously analyzed while preserving strong performance and scalability across the THz band via cavity-length tuning. \textcolor{black}{Notably, the required DC bias for lasing remains comparable to that reported for the multilayer structure in \cite{hakimi2024chiral}, despite using here only a single 2D bilayer.}

In conclusion, this work establishes BCD-driven electro-optic gain in low-symmetry 2D materials as a viable mechanism for compact, electrically driven THz amplifiers and lasers that deserve to be further investigated in an experimental manner. Beyond TBG, the proposed platform is broadly applicable to a wide class of materials with BCD, providing a general route toward frequency-tunable, polarization-selective THz sources. The combination of compactness, frequency tunability, chiral control, and cavity-enhanced operation highlights the potential of geometric electronic effects for integrated THz photonics, with promising implications for on-chip spectroscopy, sensing, and high-speed wireless communication.

\textcolor{black}{\section*{Theoretical Methods}
The semiclassical Boltzmann transport equation approach was employed to characterize the optical conductivity of a low-symmetry 2D material under a static electric bias \cite{rappoport2023engineering}. Using the conductivity matrix given in Eq. \eqref{eq:TBG}, its eigenvalues and corresponding eigenvectors (polarization eigenstates) were then determined. The transfer matrix method was employed to calculate the transmission and reflection of the proposed cavity structure shown in Fig. \ref{fig:Structure}, with the transfer matrices of the individual elements detailed in the manuscript and Supporting Information.
To determine the lasing condition, a transmission-line formalism was employed to model the cavity, consisting of dielectric layers terminated by mirrors. The resulting resonance condition, given by Eq. \eqref{eq:modes}, is generally a complex-valued equation and therefore requires a complex-valued frequency to be satisfied. For each value of the $\xi$ parameter, the exact complex frequency satisfying the resonance condition was numerically determined and plotted in Fig. \ref{fig:Complex_frequency}. The value of $\xi$ for which the resonance condition with ${\rm Im}(f)=0$ is satisfied corresponds to the onset of lasing and defines the lasing threshold, $\xi_{\rm th}$.
To obtain an approximate analytical expression for $\xi_{\rm th,a}$, algebraic manipulations and Taylor-series expansions were performed under several assumptions, including highly reflective partial mirrors, $\left|y_{\rm m}\right|\gg1$, and an interaction between the cavity field and the 2D material that results in only a slight shift of the lasing frequency from the resonance frequency of the unloaded FP cavity. Detailed derivations and additional analytical results are provided in the Supporting Information.}

\providecommand{\noopsort}[1]{}\providecommand{\singleletter}[1]{#1}%

\onecolumngrid
\newpage

\setcounter{table}{0}
\setcounter{figure}{0}
\setcounter{equation}{0}
\setcounter{section}{0}

\renewcommand{\thetable}{S\arabic{table}}
\renewcommand{\thefigure}{S\arabic{figure}}
\renewcommand{\theequation}{S\arabic{equation}}
\renewcommand{\thesection}{S\Roman{section}}
\renewcommand{\bibnumfmt}[1]{[S#1]}
\renewcommand{\citenumfont}[1]{S#1}

\makeatletter
\renewcommand{\theHfigure}{S\arabic{figure}}
\renewcommand{\theHequation}{S\arabic{equation}}
\renewcommand{\theHtable}{S\arabic{table}}
\renewcommand{\theHsection}{S\Roman{section}}
\makeatother

\begin{center}
{\bf \large{Supporting Information: \\ Chiral Terahertz Amplification and Lasing \\  using Two-Dimensional Materials with Berry Curvature Dipole}}
\end{center}

\section{Transfer Matrix Expressions}
We provide the details of the transfer matrices used to model each section of the cavity shown in Fig.~1 of the main body of the paper.
The forward transfer matrix describing wave propagation through a dielectric layer of dielectric constant $\varepsilon_{\mathrm{d}}$ and length $L/2$ represents wave propagation and is expressed as

\begin{equation}
\overline{\boldsymbol{\mathrm{T}}}_{\mathrm{d}}  =
 	\left[\begin{array}{cc}
		e^{ik_{\rm d} L/2}\overline{\mathbf{I}} & 
		0\\
		0 & 
		e^{-ik_{\rm d} L/2}\overline{\mathbf{I}}\\
	\end{array}\right],
\end{equation}
where $k_{\rm d} = (\omega/c)\sqrt{\varepsilon_{\mathrm{d}}} = k_0\sqrt{\varepsilon_{\mathrm{d}}}$ is the wavenumber in that layer. 

The transfer matrix of the partially reflective mirrors terminating the cavity on both sides, e.g., distributed Bragg reflector (DBR) mirrors, is expressed as \cite{furman2023frozen}

\begin{equation}
	\overline{\boldsymbol{\mathrm{T}}}_{\rho} = \frac{1}{\tau}
	\left[\begin{array}{cc}
		\left(\tau^2 - \rho^2\right)\overline{\mathbf{I}} & 
		\rho\overline{\mathbf{I}} \\
		
		-\rho\overline{\mathbf{I}} & 
		\overline{\mathbf{I}}\\
  
	\end{array}\right],
\end{equation}
where $\rho = \rho_0e^{i\theta_{\rho}}$, and $\tau = \sqrt{1- \rho_0^2} e^{i\left(\theta_{\rho} \pm \pi/2 \right)}$. Here, $\rho$ is the field reflection coefficient, with magnitude $\rho_0$ and phase $\theta_\rho$. Both choices of the $\pm$ sign in the definition of $\tau$ are physically valid; however, since the cavity platform shown in Fig. 1 of the paper is terminated by two identical DBR mirrors, this sign choice does not affect the final total transfer matrix of the system. \textcolor{black}{We assume that the DBRs do not incorporate any additional losses while allowing light to be partially transmitted out of the cavity in Fig. 2.} \textcolor{black}{If we remove the 2D material located at the center of the cavity, the total absorptance is zero, and the incident wave can be fully transmitted on the other side of the resonant cavity. When the 2D material is present and when gain overcomes the material losses, transmittance can be larger than unity at resonance.}

\section{Optimal Amplification Operating Point for Cavity Design}
We aim to determine the optimal cavity length that maximizes amplification of the transmitted field as the frequency of the illuminating incident wave is varied, and for different values of the loss parameter $\gamma$. In addition, practical applications may require operation at different frequencies, motivating the need for a scalable cavity design.
To address this, we calculate the transmittance of the cavity shown in Fig. 1 of the main body of the paper, under RCP illumination from the left, as a function of both the cavity length $L$ and the illuminating frequency using Eq. (15) of the paper.

\begin{figure}[h]
    \centering
    \includegraphics[width=0.6\linewidth]{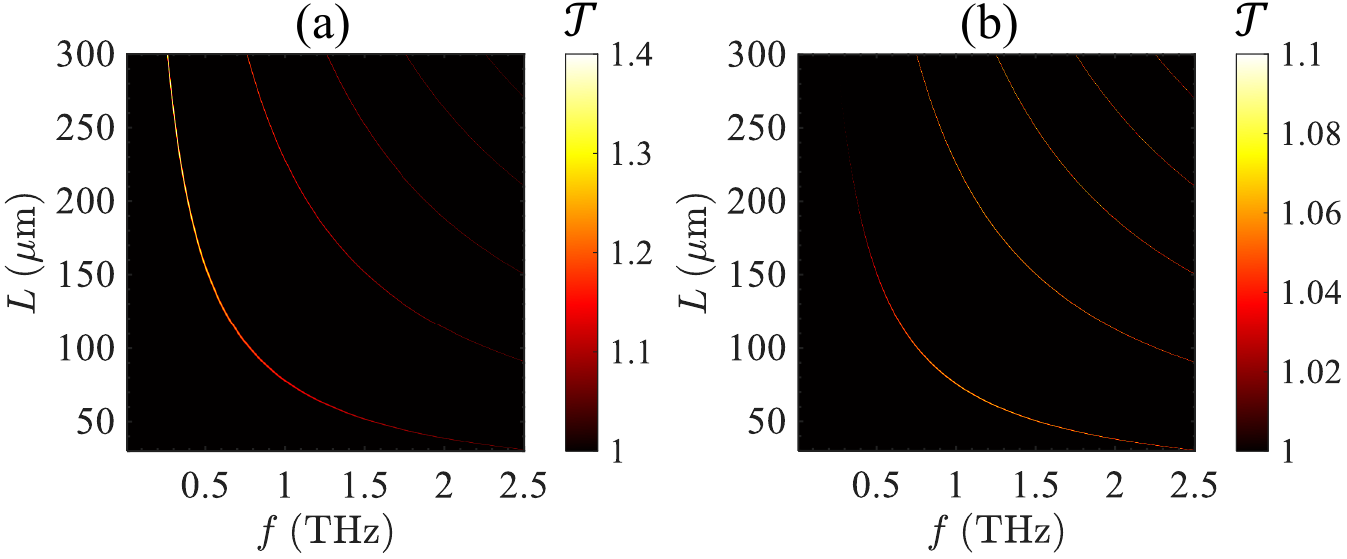}
    \caption{Transmittance under RCP illumination from the left by varying cavity length and illuminating frequency for (a) $\gamma = 10^{12} \: {\rm s^{-1}}$, (b) $\gamma = 5 \times 10^{12} \: {\rm s^{-1}}$. The remaining parameters are $\varepsilon_{\rm d} = 4$, $\omega_{\rm F}/(2\pi) = 0.24 \: {\rm THz}$, $\xi/\omega_{\rm F} = 5$, and $\rho = -0.9$.}
    \label{fig:Optimal_amplification_Supp}
\end{figure}

Figure~\ref{fig:Optimal_amplification_Supp} presents the resulting transmittance maps for different values of $\gamma$. The contour lines indicate the combinations of frequency and cavity length that yield a transmittance greater than unity, with brighter contours corresponding to higher transmittance.
We emphasize that this cavity design is general and can be applied to any 2D material exhibiting a non-Hermitian electro-optic response associated with a Berry curvature dipole that leads to gain and amplification, allowing similar optimization to identify the optimal operating point.

\section{Polarization of Transmitted Waves Through the Cavity}
We analyze the polarization of the field transmitted through the cavity shown in Fig. 1 of the main body of the paper. As explained there, the cavity is illuminated from the left with circularly polarized (CP) incident waves, and depending on the handedness of the incident wave, either the right- or left-circularly polarized (RCP and LCP) can exhibit amplified transmittance, which can be controlled by the direction of the applied DC bias. To this end, the transmitted electric field is first expressed as $\mathbf{E}_{\rm t} = \overline{\mathbf{t}} \: \mathbf{E}_{\rm inc}$, where $\overline{\mathbf{t}}$ is the transmission matrix of the cavity, defined in Eq. (14) of the paper. The transmitted electric field is then projected onto the RCP and LCP basis to extract the co- and cross-polarized components, yielding


\begin{equation}
    E_{\rm t}^{\rm co-pol, \: \pm} = \left|\mathbf{e}_{\pm}^\dagger \: \overline{\mathbf{t}} \: \mathbf{e}_{\pm}\right|, \quad \quad
    E_{\rm t}^{\rm x-pol, \: \pm} = \left|\mathbf{e}_{\mp}^\dagger \: \overline{\mathbf{t}} \: \mathbf{e}_{\pm}\right|,
\end{equation}
where the $\dagger$ denotes transpose conjugation. Here, $(+)$ and $(-)$ correspond to RCP and LCP, respectively, and $\mathbf{e}_{\pm} = \frac{1}{\sqrt{2}} \left[ 1, \,  \pm i \right]^{\rm T}$ denote the unit vectors associated with RCP $(+)$ and LCP $(-)$.

\begin{figure}[b]
    \centering
    \includegraphics[width=0.6\linewidth]{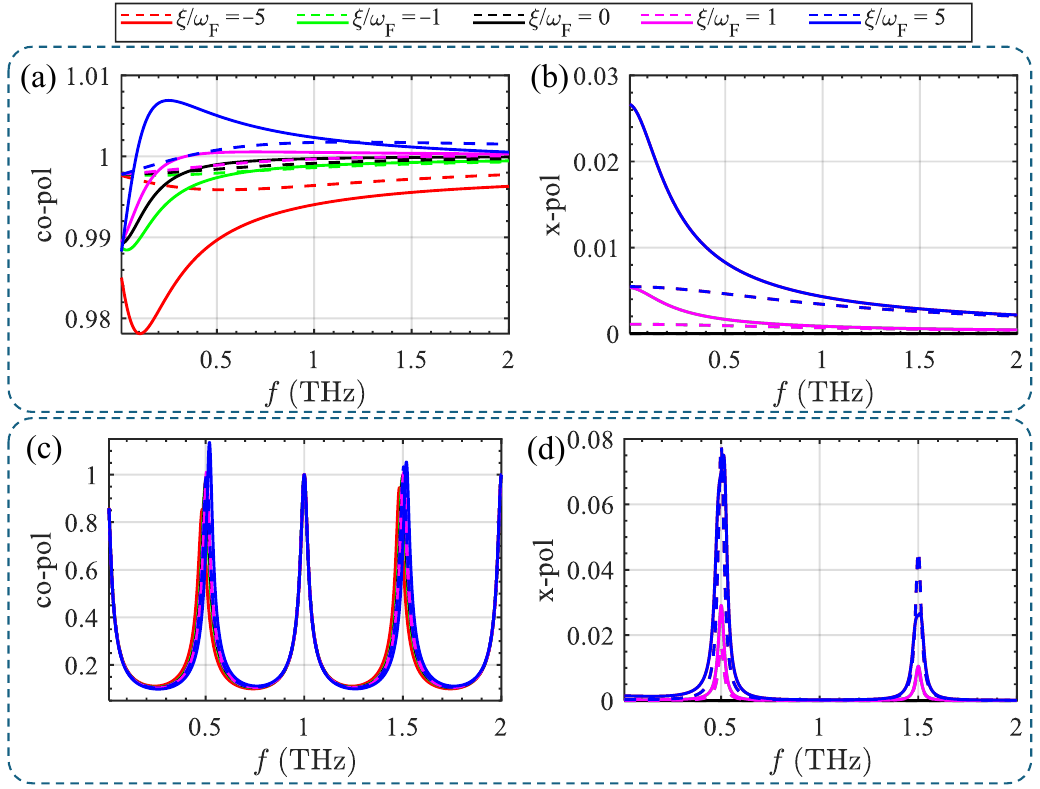}
    \caption{Transmitted co- and cross-polarization components for RCP wave incidence on the TBG without (a,b) and with (c,d) cavity versus frequency for different values of the gain parameter $\xi$. Solid and dashed curves correspond to $\gamma = 10^{12} \: {\rm s^{-1}}$ and $\gamma = 5 \times 10^{12} \: {\rm s^{-1}}$, respectively. The other parameters are $\varepsilon_{\rm d} = 4$, $\omega_{\rm F}/(2\pi) = 0.24 \: {\rm THz}$. In (c,d), the cavity with TBG at its center has $L = 149.90\: \mu{\rm m}$ and $\rho = -0.9$. The incident field is coming from the left (see Fig. 1 of the main body of the paper); if it came from the right, amplification would occur for negative values of the gain parameter $\xi$.}
    \label{fig:Polarizations_Supp}
\end{figure}

Figure \ref{fig:Polarizations_Supp} shows the co- and cross-polarization components of the transmitted field under RCP illumination for (top row) a twisted-bilayer graphene (TBG) {\em without} cavity, and (bottom row) a cavity with TBG at its center, for different values of the gain parameter $\xi$. Solid and dashed curves correspond to $\gamma = 10^{12}\,\mathrm{s^{-1}}$ and $\gamma = 5\times10^{12}\,\mathrm{s^{-1}}$, respectively.
In the top row, the cross-polarization component is negligible compared to the co-polarization component, indicating that the polarization of the incident circularly polarized wave is largely preserved upon transmission through the single TBG, regardless of variations in the material parameters.
In the bottom row, the co- and cross-polarization components for the designed cavity are shown. Here, the cross-polarization contribution is slightly increased due to the increased light-matter interaction provided by the cavity, which amplifies the intrinsic response of the 2D material. Nevertheless, the co-polarization component remains dominant, indicating that the transmitted wave retains the same polarization as the incident wave while experiencing significantly stronger amplification. Therefore, the proposed cavity design substantially enhances amplification without altering the polarization state of the incident wave.  

In Fig.~\ref{fig:Polarizations_Supp}, the incident field is coming from the left (see Fig. 1 of the main body of the paper) with RCP polarization. For an LCP polarization incident from the left side, amplification would be observed for negative values of the gain parameter $\xi$. Since the same behavior would be observed for an LCP incident wave (except for a flip in the sign of $\xi$), these numerical results are omitted for brevity. Furthermore, if RCP is illuminating the TBG from the right side, amplification would occur for negative values of the gain parameter $\xi$ due to the chiral properties of TBG.

\section{Derivation of Resonance Condition using Impedance Concept}
\label{sec:Resonance_Condition_Supp}

\begin{figure}[h]
    \centering
    \includegraphics[width=0.5\linewidth]{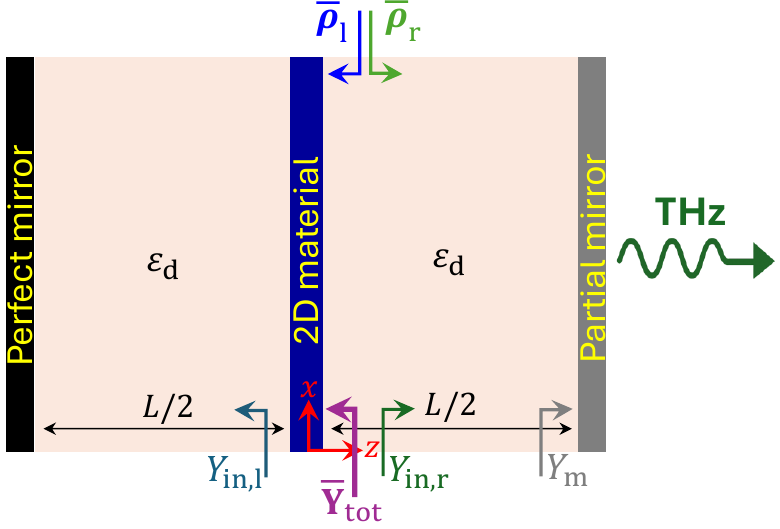}
    \caption{Fabry-Pérot cavity with a 2D material exhibiting BCD positioned at the center to enhance light-matter interaction, terminated by a perfect mirror on one side and a partial mirror on the other to achieve THz lasing.}
    \label{fig:Cavity_Supp}
\end{figure}

We determine the reflection coefficient, which is essential for identifying the resonance condition in the cavity shown in Fig. \ref{fig:Cavity_Supp}. To this end, we model the structure using transmission-line theory, a powerful tool for describing a large class of wave propagation and finding the reflection coefficient based on the impedance mismatch between different media.

In this framework, the mirrors are represented as loads, and the 2D material at the center of the cavity is modeled as a shunt admittance. To obtain $\overline{\boldsymbol{\rho}}_{\mathrm{r}}$ and $\overline{\boldsymbol{\rho}}_{\mathrm{l}}$, we first calculate the admittances shown in Fig. \ref{fig:Cavity_Supp}. The admittance seen from the right, $Y_{\mathrm{in,r}}$, corresponding to a dielectric layer of length $L/2$ terminated by a partially reflective mirror is given by \cite{pozar2009microwave}

\begin{equation}
    Y_{\mathrm{in,r}} = Y_\mathrm{d} \frac{Y_{\mathrm{m}} - iY_{\mathrm{d}}\tan(k_{\mathrm{d}}L/2)}{Y_{\mathrm{d}} - iY_{\mathrm{m}}\tan(k_{\mathrm{d}}L/2)},
\label{eq:Y_in_l_Supp}
\end{equation}
where $Y_{\mathrm{d}} = Y_0  \sqrt{\varepsilon_{\rm d}} = \sqrt{\varepsilon_0 \varepsilon_{\rm d} / \mu_0}$ is the characteristic admittance of the dielectric layer, and $Y_{\mathrm{m}}$ is the admittance of the partially reflective mirror. For the partially reflective mirror on the right, the field-reflection coefficient $\rho$ is provided by $\rho = \left(Y_{\mathrm{d}} - Y_{\mathrm{m}}\right)/\left(Y_{\mathrm{d}} + Y_{\mathrm{m}}\right)$. Accordingly, the corresponding load admittance $Y_{\mathrm{m}}$ (i.e., the admittance of the partially reflective mirror) is determined as

\begin{equation}
    Y_{\mathrm{m}} = Y_{\mathrm{d}} \frac{1-\rho}{1+\rho} = y_{\mathrm{m}} Y_{\mathrm{d}},
\label{eq:yL_Supp}
\end{equation}
where $y_{\mathrm{m}} = Y_{\mathrm{m}}/Y_{\mathrm{d}} = (1-\rho)/(1+\rho)$ denotes the normalized admittance of the partially reflective mirror seen from the cavity. Substituting this into Eq. \eqref{eq:Y_in_l_Supp} yields

\begin{equation}
    Y_{\mathrm{in,r}} = Y_\mathrm{d} \frac{y_{\mathrm{m}} - i\tan(k_{\mathrm{d}}L/2)}{1 - iy_{\mathrm{m}} \tan(k_{\mathrm{d}}L/2)}.
\end{equation}

On the left side of the 2D material, the admittance $Y_{\mathrm{in,l}}$ is seen toward the left of the dielectric layer of length $L/2$ terminated by a perfect (fully reflective) mirror. Here, we consider the perfect mirror as a perfect electric conductor (PEC), i.e., a short circuit. Accordingly, the input admittance in configuration $Y_{\mathrm{in,l}}$ is given by \cite{pozar2009microwave}

\begin{equation}
    Y_{\mathrm{in,l}} = iY_{\mathrm{d}} \cot(k_{\mathrm{d}} L/2).
\end{equation}
The 2D material is placed in parallel with $Y_{\mathrm{in,l}}$, as shown in Fig. \ref{fig:Cavity_Supp}. Thus, the total admittance $\overline{\textbf{Y}}_{\mathrm{tot}}$ is

\begin{equation}
    \overline{\textbf{Y}}_{\mathrm{tot}} = Y_{\mathrm{in,l}} \overline{\textbf{I}} + \overline{\boldsymbol{\sigma}},
\end{equation}
where $\overline{\boldsymbol{\sigma}}$ is the conductivity tensor of the TBG given in Eq. (1) of the paper. Then, using the admittances derived above, the reflection coefficients $\overline{\boldsymbol{\rho}}_{\mathrm{l}}$ and $\overline{\boldsymbol{\rho}}_{\mathrm{r}}$ are expressed as

\begin{gather}
    \overline{\boldsymbol{\rho}}_{\mathrm{l}} = \left(\overline{\textbf{Y}}_{\mathrm{d}} + \overline{\textbf{Y}}_{\mathrm{tot}}\right)^{-1} \left(\overline{\textbf{Y}}_{\mathrm{d}} - \overline{\textbf{Y}}_{\mathrm{tot}}\right), \label{eq:rho_l_supp} \\
    \overline{\boldsymbol{\rho}}_{\mathrm{r}} = \left(\overline{\textbf{Y}}_{\mathrm{d}} + \overline{\textbf{Y}}_{\mathrm{in,r}}\right)^{-1} \left(\overline{\textbf{Y}}_{\mathrm{d}} - \overline{\textbf{Y}}_{\mathrm{in,r}}\right),
\end{gather}
with $\overline{\textbf{Y}}_{\mathrm{in,r}} = Y_{\mathrm{in,r}}\overline{\textbf{I}}$, and $\overline{\textbf{Y}}_{\mathrm{d}} = Y_{\mathrm{d}}\overline{\textbf{I}}$. Then, $\overline{\boldsymbol{\rho}}_{\mathrm{r}}$ is simplified as,

\begin{equation}
    \overline{\boldsymbol{\rho}}_{\mathrm{r}} = \frac{Y_{\mathrm{d}} - Y_{\mathrm{in,r}}}{Y_{\mathrm{d}} + Y_{\mathrm{in,r}}}\overline{\textbf{I}} = \frac{1 - y_{\mathrm{m}}}{1 + y_{\mathrm{m}}}\exp(i k_{\mathrm{d}} L)\overline{\textbf{I}} = \rho_{\rm r}\overline{\textbf{I}},
    \label{eq:rho_r_supp}
\end{equation}
where $\rho_{\rm r} = \rho \exp(i k_{\mathrm{d}} L) =  \left(1 - y_{\mathrm{m}}\right)\exp(i k_{\mathrm{d}} L)/\left(1 + y_{\mathrm{m}}\right)$. With these definitions, the resonance condition obtained in the paper, $F(\omega) \equiv \det \left(\overline{\boldsymbol{\rho}}_{\mathrm{r}} \overline{\boldsymbol{\rho}}_{\mathrm{l}} - \overline{\textbf{I}} \right) = 0$, simplifies to

\begin{equation}
    \frac{\left[\left(1+ \rho_{\rm r} \right) \sigma_{xx} + \left(Y_+ - \rho_{\rm r} Y_- \right) \right]^2 - (1+\rho_{\rm r})^2\sigma_{xy}\sigma_{yx}}{(\sigma_{xx} + Y_+)^2 - \sigma_{xy}\sigma_{yx}} = 0,
\label{eq:resonance_Supp}
\end{equation}
where

\begin{gather}
\begin{split}
    Y_+ = Y_{\mathrm{d}} + Y_{\rm in,l} = Y_{\mathrm{d}} \left(1+ i\cot \left(k_{\mathrm{d}} L/2 \right) \right) = i Y_{\mathrm{d}} \csc \left(k_{\mathrm{d}} L/2 \right) \exp \left( -i k_{\mathrm{d}} L/2 \right), \label{eq:Y_p_Supp} \end{split} \\
    \begin{split}
    Y_- = Y_{\mathrm{d}} - Y_{\rm in,l} = Y_{\mathrm{d}} \left(1 - i\cot \left(k_{\mathrm{d}} L/2 \right) \right) = -i Y_{\mathrm{d}} \csc \left(k_{\mathrm{d}} L/2 \right) \exp \left( i k_{\mathrm{d}} L/2 \right). \label{eq:Y_m_Supp}
    \end{split}
\end{gather}
The complex-frequency resonance condition is satisfied either when the numerator of Eq. \eqref{eq:resonance_Supp} becomes zero or when its denominator diverges to infinity. \textcolor{black}{As indicated by Eqs. \eqref{eq:Y_p_Supp} and \eqref{eq:Y_m_Supp}, both the numerator and denominator exhibit poles due to the presence of the $\csc(\cdot)$ function. However, when $\rho=-1$ (lossless cavity), a pole-zero cancellation occurs at the numerator, resulting in a resonance condition associated with a pole at the denominator arising from the $\csc(k_{\rm d}L/2)$ function in Eq. \eqref{eq:resonance_Supp}, corresponding to antisymmetric resonance-mode profiles with an electric field node at the cavity center. Therefore, the solutions of $\sin(k_{\rm d}L/2) = 0$ are related to the even-order resonances of the FP cavity discussed in the paper.} 

\textcolor{black}{Instead, when the partial mirror is not highly reflective, i.e., in the realistic case of $\rho\neq-1$, corresponding to a {\em partially} reflective mirror, the poles at the numerator and denominator cancel each other. Therefore, a pole of $\csc(k_{\rm d}L/2)$ at the denominator does not constitute a root of Eq. \eqref{eq:resonance_Supp} anymore. The exact resonance condition at a complex-valued frequency for these modes deviates slightly from that predicted by $\sin(k_{\rm d}L/2) = 0$. In this paper, we do not consider this resonance condition since our focus is on the odd-order resonances with a maximum field at the center to establish a strong light-matter interaction.} 

We now focus on the resonances arising from the zeros of the numerator 
\begin{equation}
    N(\omega) \equiv \sigma_{xx} \mp \sqrt{\sigma_{xy}\sigma_{yx}}+ \frac{Y_+ - \rho_{\rm r} Y_-}{1+ \rho_{\rm r}} = 0,
\label{eq:mode2_Supp}
\end{equation}
which correspond to the odd-order resonances, i.e., those with a maximum of electric field at the center. This relation is further simplified using Eqs. \eqref{eq:Y_p_Supp} and \eqref{eq:Y_m_Supp}. Using $Y_{\rm d} = 1/\eta_{\rm d}$, the final condition for the odd-order FP resonances becomes

\begin{equation}
    N(\omega) \equiv \sigma_{xx} \mp  \sqrt{\sigma_{xy}\sigma_{yx}}  +i \frac{1}{\eta_{\mathrm{d}}} \frac{1 -  \cot^2(k_{\mathrm{d}} L/2) +2i y_{\mathrm{m}} \cot(k_{\mathrm{d}} L/2) }{iy_{\mathrm{m}} - \cot(k_{\mathrm{d}} L/2)} = 0.
\label{eq:modes2_Supp}
\end{equation}

\section{Polarization Eigenstates of the Lasing Cavity}

We now determine the polarization eigenstates of the modes inside the lasing cavity. Based on Eq. (17) of the main body of the paper, the resonance condition is given by, $\overline{\boldsymbol{\rho}}_{\mathrm{r}} \overline{\boldsymbol{\rho}}_{\mathrm{l}} \mathbf{E} = \mathbf{E}$. Defining $\overline{\mathbf{M}} = \overline{\boldsymbol{\rho}}_{\mathrm{r}} \overline{\boldsymbol{\rho}}_{\mathrm{l}}$, the resonance condition is rewritten as

\begin{equation}
    \overline{\mathbf{M}} \mathbf{E} = \mathbf{E},
    \label{eq:lasing_supp}
\end{equation}
Here, Eq. \eqref{eq:lasing_supp} is an eigenvalue problem in which the eigenvalue is unity. The eigenvectors of $\overline{\mathbf{M}}$ therefore represent the polarization states of the lasing cavity.
Using the definition of $\overline{\boldsymbol{\rho}}_{\mathrm{l}}$ in Eq. \eqref{eq:rho_l_supp} and $Y_{\pm}$ in Eqs. \eqref{eq:Y_m_Supp} and \eqref{eq:Y_p_Supp}, $\overline{\boldsymbol{\rho}}_{\mathrm{l}}$ is expressed as

\begin{equation}
    \overline{\boldsymbol{\rho}}_{\mathrm{l}} = \left(Y_+ \overline{\textbf{I}} + \overline{\boldsymbol{\sigma}} \right)^{-1} \left(Y_- \overline{\textbf{I}} - \overline{\boldsymbol{\sigma}} \right).
\end{equation}
After some algebraic simplification, this becomes

\begin{equation}
    \overline{\boldsymbol{\rho}}_{\mathrm{l}} = \frac{1}{(Y_+ + \sigma_{xx})^2 - \sigma_{xy}\sigma_{yx}} \left[ \left( Y_+ Y_-  - \det\left(\overline{\boldsymbol{\sigma}} \right) + 2\sigma_{xx}Y_- \right) \overline{\mathbf{I}} - 2Y_{\rm d} \overline{\boldsymbol{\sigma}} \right].
\end{equation}
Using the definition of $\overline{\boldsymbol{\rho}}_{\mathrm{r}}$ from Eq. \eqref{eq:rho_r_supp}, the final expression for $\overline{\mathbf{M}}$ reads

\begin{equation}
    \overline{\mathbf{M}} = \frac{\rho_{\rm r}}{(Y_+ + \sigma_{xx})^2 - \sigma_{xy}\sigma_{yx}} \left[ \left( Y_+ Y_-  - \det\left(\overline{\boldsymbol{\sigma}} \right) + 2\sigma_{xx}Y_- \right) \overline{\mathbf{I}} - 2Y_{\rm d} \overline{\boldsymbol{\sigma}} \right].
\label{eq:M_supp}
\end{equation}
This is expressed in the compact form

\begin{equation}
    \overline{\mathbf{M}} = \alpha \overline{\mathbf{I}} + \beta \overline{\boldsymbol{\sigma}} = 
    \begin{bmatrix} 
    \alpha + \beta \sigma_{xx} & \beta \sigma_{xy} \\ 
    \beta \sigma_{yx} & \alpha + \beta \sigma_{xx} 
    \end{bmatrix},
\end{equation}
where $\alpha$ and $\beta$ are determined from Eq. \eqref{eq:M_supp}. Then, the eigenvalues and eigenvectors of $\overline{\mathbf{M}}$ are obtained: 
\begin{subequations}
\begin{equation}
    \lambda_{1,2} = \alpha + \beta \left( \sigma_{xx} \mp  \sqrt{\sigma_{xy}\sigma_{yx}} \right),
\label{eq:eigenval_Supp}
\end{equation}
\begin{equation}
    \mathbf{V}_{1,2} = \begin{bmatrix} \mp \sqrt{\frac{\sigma_{xy}}{\sigma_{yx}}} \\ 1 \end{bmatrix}.
\end{equation}
\label{eq:eigenvec_Supp}
\end{subequations}
Imposing the condition $\lambda_{1,2}=1$ in Eq. \eqref{eq:eigenval_Supp} and performing straightforward simplifications leads to the same resonance equation obtained in Eq. \eqref{eq:mode2_Supp}.
Hence, we have proven that the polarization states of the lasing cavity are identical to the eigenvectors of $\overline{\boldsymbol{\sigma}}$ given in Eq. (3) of the paper. This also demonstrates that the $-$ and $+$ signs in the resonance condition given in Eq. \eqref{eq:modes2_Supp} correspond to the polarization eigenstates of the 2D material $\mathbf{E}^{\sigma}_1$ and $\mathbf{E}^{\sigma}_2$ in Eq. (3) of the main text, respectively. Under the lasing condition, which requires amplification of one of the two modes of the 2D material, the mode $\mathbf{E}_1^\sigma$ propagates and amplifies inside the cavity when $\xi>0$ is above a certain threshold value.

\section{Derivation of Lasing Threshold}
\label{sec:Lasing_Threshold_Supp}
We aim to determine the threshold value of the gain parameter ($\xi$) corresponding to the onset of lasing. We start from the dispersion equation for the resonances interacting with the 2D material (odd-order FP resonances), given in Eq. \eqref{eq:modes2_Supp},

\begin{equation}
        \sigma_{xx} \mp  \sqrt{\sigma_{xy}\sigma_{yx}} + i \frac{1}{\eta_{\mathrm{d}}} \frac{1 -  \cot^2(k_{\mathrm{d}} L/2)  +2i y_{\mathrm{m}} \cot(k_{\mathrm{d}} L/2) }{iy_{\mathrm{m}} - \cot(k_{\mathrm{d}} L/2)}= 0.
\end{equation}
Assuming that the cavity resonance frequency is close to that of the unloaded FP cavity and that the frequency shift due to interaction with the 2D material is small, we have \textcolor{black}{$k_{\mathrm{d}} L/2 \approx q\pi/2$ with $q=1,3,5,\dots$}, which implies $\left| \cot(k_{\mathrm{d}} L/2) \right| \ll 1$, where $q$ is the FP resonance index. Since this dispersion equation corresponds to odd-order resonances, only odd integers are considered. In addition, the partial mirror on the right side of the cavity (Fig.~\ref{fig:Cavity_Supp}) is assumed to have high reflectivity with a constant phase of $\pi$, i.e., a reflection coefficient close to $-1$, which leads to $\left|y_{\mathrm{m}} \right| \gg 1$. Under these approximations, the dispersion equation is simplified as

\begin{equation}
        \sigma_{xx} \mp  \sqrt{\sigma_{xy}\sigma_{yx}} + i \frac{1}{\eta_{\mathrm{d}}} \frac{1   +2i y_{\mathrm{m}} \cot(k_{\mathrm{d}} L/2) }{iy_{\mathrm{m}}} \approx 0.
\end{equation}
After straightforward algebraic manipulation and using the trigonometric identity to express $\cot(\cdot)$ in terms of $\tan(\cdot)$, the dispersion relation reduces to

\begin{equation}
    \sigma_{xx} \mp  \sqrt{\sigma_{xy}\sigma_{yx}} + i \frac{1}{\eta_{\mathrm{d}}} \left(-i y_{\mathrm{m}}^{-1} + 2 \tan \left(\frac{q\pi}{2} - \frac{k_{\mathrm{d}} L}{2} \right) \right) \approx 0.
\end{equation}
Assuming the cavity resonance remains close to that of the FP cavity without TBG, the argument of the $\tan(\cdot)$ function is close to zero, allowing us to use the first-order Taylor expansion of $\tan(\cdot)$. Under this approximation, the dispersion relation becomes
  
\begin{equation}
    \sigma_{xx} \mp  \sqrt{\sigma_{xy}\sigma_{yx}} \approx i \frac{1}{\eta_{\mathrm{d}}} \left(i y_{\mathrm{m}}^{-1} + (k_{\mathrm{d}} L - q\pi) \right).
\label{eq:reonance_approx_Supp}
\end{equation}
The left-hand side of the above equation corresponds to the eigenvalues $\sigma_{1,2}$ of the conductivity matrix $\overline{\boldsymbol{\sigma}}$, given in Eq. (2) of the paper, associated with the polarization eigenstates $\mathbf{E}^{\sigma}_{1,2}$ in Eq. (3) of the paper. Since we are interested in the lasing threshold, only the polarization eigenstate that exhibits gain (polarization eigenstate 1, when $\xi>0$) is relevant; therefore, we consider only $\sigma_1$ that has the negative sign. Conversely, when $\xi<0$, polarization eigenstate 2 is the one that experiences gain. 

These eigenvalues are approximated in two distinct regimes, determined by the comparison between the operating angular frequency $\omega$ and losses $\gamma$. In particular, we consider: (i) the high-frequency, low-loss regime ($\omega \gg \gamma$), and (ii) the low-frequency, high-loss regime ($\omega \ll \gamma$). In each regime, we employ a Taylor series expansion to approximate the eigenvalue of the conductivity matrix, which simplifies the left-hand side of the above equation.
First, when $\omega \gg \gamma$, we obtain

\begin{equation}
    \sigma_{xx} - \sqrt{\sigma_{xy}\sigma_{yx}} \approx i\sigma_0 \left[\left( \frac{\omega_{\rm F}}{\omega} + \frac{\xi}{\gamma} \right) + i\frac{\gamma}{\omega} \left( -\frac{\omega_{\rm F}}{\omega} + \frac{\xi}{2\gamma} \right) + \left(\frac{\gamma}{\omega}\right)^2 \left( -\frac{\omega_{\rm F}}{\omega} + \frac{5\xi}{8\gamma} \right) +  i\left(\frac{\gamma}{\omega}\right)^3 \left(\frac{\omega_{\rm F}}{\omega} - \frac{13\xi}{16\gamma} \right)\right].
\end{equation}
Substituting this approximation into Eq.~\eqref{eq:reonance_approx_Supp} yields

\begin{equation}
    i\sigma_0 \left[\left( \frac{\omega_{\rm F}}{\omega} + \frac{\xi}{\gamma} \right) + i\frac{\gamma}{\omega} \left( -\frac{\omega_{\rm F}}{\omega} + \frac{\xi}{2\gamma} \right) + \left(\frac{\gamma}{\omega}\right)^2 \left( -\frac{\omega_{\rm F}}{\omega} + \frac{5\xi}{8\gamma} \right) +  i\left(\frac{\gamma}{\omega}\right)^3 \left(\frac{\omega_{\rm F}}{\omega} - \frac{13\xi}{16\gamma} \right)\right] \approx i \frac{1}{\eta_{\mathrm{d}}} \left(i y_{\mathrm{m}}^{-1} + (k_{\mathrm{d}} L - q\pi) \right).
\end{equation}
Rewriting this equation by separating the real and imaginary parts, we obtain

\begin{equation}
\begin{split}
    \left[\eta_{\mathrm{d}}\sigma_0 \left(\left( \frac{\omega_{\rm F}}{\omega} + \frac{\xi}{\gamma} \right) + \left(\frac{\gamma}{\omega}\right)^2 \left( -\frac{\omega_{\rm F}}{\omega} + \frac{5\xi}{8\gamma} \right) \right) + q\pi - \frac{\omega}{2f_1} \right] + \\ i\left[ \eta_{\mathrm{d}}\sigma_0 \left( \frac{\gamma}{\omega} \left( -\frac{\omega_{\rm F}}{\omega} + \frac{\xi}{2\gamma} \right) + \left(\frac{\gamma}{\omega}\right)^3 \left(\frac{\omega_{\rm F}}{\omega} - \frac{13\xi}{16\gamma} \right)\right) - y_{\mathrm{m}}^{-1} \right] \approx 0.
    \label{eq:Separated_Resonance_Eq_Supp}
\end{split}
\end{equation}
By solving this complex-valued equation for $\omega$, we can determine the complex lasing frequency and the corresponding required gain parameter. Under the time-evolution convention $e^{-i\omega t}$ adopted here, lasing occurs when ${\rm Im}(\omega) > 0$. Therefore, the lasing threshold corresponds to the point at which the complex frequency crosses the real axis in the complex-$\omega$ plane, indicating the onset of lasing.
Both the real and imaginary parts of Eq.~\eqref{eq:Separated_Resonance_Eq_Supp} must simultaneously vanish at this crossing, resulting in a coupled system of two equations for real $\omega$ and $\xi$. The real part is primarily important to determine the lasing frequency, which remains close to the unloaded FP cavity resonance because $\eta_{\mathrm{d}}\sigma_0 \ll 1$. Consequently, the lasing threshold condition is well approximated by requiring only the imaginary part of Eq.~\eqref{eq:Separated_Resonance_Eq_Supp} to vanish: 

\begin{equation}
    \eta_{\mathrm{d}}\sigma_0 \left( \frac{\gamma}{\omega} \left( -\frac{\omega_{\rm F}}{\omega} + \frac{\xi}{2\gamma} \right) + \left(\frac{\gamma}{\omega}\right)^3 \left(\frac{\omega_{\rm F}}{\omega} - \frac{13\xi}{16\gamma} \right)\right) - y_{\mathrm{m}}^{-1} \approx 0.
\end{equation}
Here, $\omega$ must be specified to determine the corresponding value of $\xi$. We take \textcolor{black}{$\omega \approx 2\pi q f_1 \equiv \omega_{\rm r,a}$}, which is the approximate resonance angular frequency, assumed equal to that of the TBG-unloaded FP cavity.
Finally, the corresponding approximate gain parameter $\xi_{\rm th,a}$ required for the onset of lasing when $\omega \gg \gamma$ is obtained as

\begin{equation}
    \xi_{\rm th,a} ({\omega \gg \gamma}) \approx 2\left(\frac{\omega_{\rm r,a}}{y_{\mathrm{m}}\eta_{\mathrm{d}}\sigma_0} + \gamma\frac{\omega_{\rm F}}{\omega_{\rm r,a}}\left(1 - \left(\frac{\gamma}{\omega_{\rm r,a}}\right)^2\right)\right) \left(1 - \frac{13}{8}\left(\frac{\gamma}{\omega_{\rm r,a}}\right)^2\right)^{-1}.
\end{equation}
Then, the same analysis is carried out for $\omega \ll \gamma$. In this case, the approximate eigenvalue of the conductivity is obtained as

\begin{equation}
    \sigma_{xx} -  \sqrt{\sigma_{xy}\sigma_{yx}} \approx i\sigma_0 \left[\left( -i\frac{\omega_{\rm F}}{\gamma} +  \frac{\sqrt{2}\xi}{\gamma} \right) + \frac{\omega}{\gamma} \left( \frac{\omega_{\rm F}}{\gamma} + i\frac{\sqrt{2}\xi}{4\gamma} \right) + \left(\frac{\omega}{\gamma}\right)^2 \left( i\frac{\omega_{\rm F}}{\gamma} - \frac{7\sqrt{2}\xi}{32\gamma} \right) +  \left(\frac{\omega}{\gamma}\right)^3 \left( -\frac{\omega_{\rm F}}{\gamma} - i\frac{25\sqrt{2}\xi}{128\gamma} \right) \right].
\end{equation}
Following the same procedure as in the previous case, the corresponding approximate gain parameter $\xi_{\rm th,a}$ required for the onset of lasing when $\omega \ll \gamma$ is obtained as




\begin{equation}
    \xi_{\rm th,a} ({\omega \ll \gamma}) \approx 2\sqrt{2} \frac{\gamma}{\omega_{\rm r,a}}\left(\frac{\gamma}{y_{\mathrm{m}}\eta_{\mathrm{d}}\sigma_0} + \omega_{\rm F}\left(1 - \left(\frac{\omega_{\rm r,a}}{\gamma}\right)^2\right)\right) \left(1 - \frac{25}{32}\left(\frac{\omega_{\rm r,a}}{\gamma}\right)^2\right)^{-1}.
\end{equation}

\providecommand{\noopsort}[1]{}\providecommand{\singleletter}[1]{#1}%
%





\end{document}